\documentclass[prd,a4paper,twocolumn,nofootinbib,showkeys]{revtex4}

\usepackage[frenchb,english]{babel}
\usepackage[latin1]{inputenc}
\usepackage{enumerate}
\usepackage{amsmath}
\usepackage{amsthm}
\usepackage{amssymb}
\usepackage{graphicx}
\usepackage{floatflt}
\usepackage{fancybox}
\usepackage{appendix}
\usepackage{enumitem}
\usepackage{subfigure}
\usepackage{array}

\usepackage{hyperref}
\usepackage[b]{esvect}

\usepackage{color}

\newcommand{\lam}{\lambda}
\newcommand{\Gam}{\Gamma}
\newcommand{\gam}{\gamma}

\newcommand{\varep}{\varepsilon}

\renewcommand{\d}{{\rm d}}

\newcommand{\p}{\partial}

\newcommand{\be} {\begin{equation}}
\newcommand{\ee} {\end{equation}}
\newcommand{\bsub}{\begin{subequations}}
\newcommand{\esub}{\end{subequations}}
\newcommand{\bea}{\begin{eqnarray}}
\newcommand{\eea}{\end{eqnarray}}
\newcommand{\bi} {\begin{itemize}}
\newcommand{\ei} {\end{itemize}}
\newcommand{\ben} {\begin{enumerate}}
\newcommand{\een} {\end{enumerate}}
\newcommand{\bmat} {\begin{pmatrix}}
\newcommand{\emat} {\end{pmatrix}}
\newcommand{\bal} {\begin{aligned}}
\newcommand{\eal} {\end{aligned}}
\newcommand{\btab}{\begin{tabular}}
\newcommand{\etab}{\end{tabular}}

\newcommand{\eq}[1]{equation~\eqref{#1}}

\usepackage{ulem}

\begin{document}
\selectlanguage{english}

\title{Robustness of topological corner modes against disorder  

and application to acoustic networks}

\author{Antonin Coutant}
\email{antonin.coutant@univ-lemans.fr}
\affiliation{Laboratoire d'Acoustique de l'Université du Mans, Unite Mixte de Recherche 6613, Centre National de la Recherche Scientifique, Avenue O. Messiaen, F-72085 Le Mans Cedex 9, France}
\affiliation{Institut de Math\' ematiques de Bourgogne (IMB), UMR 5584, CNRS, Universit\' e de Bourgogne Franche-Comt\' e, F-21000 Dijon, France}

\author{Vassos Achilleos} 
\email{achilleos.vassos@univ-lemans.fr}
\affiliation{Laboratoire d'Acoustique de l'Université du Mans, Unite Mixte de Recherche 6613, Centre National de la Recherche Scientifique, Avenue O. Messiaen, F-72085 Le Mans Cedex 9, France}

\author{Olivier Richoux}
\email{Olivier.Richoux@univ-lemans.fr}
\affiliation{Laboratoire d'Acoustique de l'Université du Mans, Unite Mixte de Recherche 6613, Centre National de la Recherche Scientifique, Avenue O. Messiaen, F-72085 Le Mans Cedex 9, France}

\author{Georgios Theocharis}
\email{georgios.theocharis@univ-lemans.fr}
\affiliation{Laboratoire d'Acoustique de l'Université du Mans, Unite Mixte de Recherche 6613, Centre National de la Recherche Scientifique, Avenue O. Messiaen, F-72085 Le Mans Cedex 9, France}

\author{Vincent Pagneux}
\email{vincent.pagneux@univ-lemans.fr}
\affiliation{Laboratoire d'Acoustique de l'Université du Mans, Unite Mixte de Recherche 6613, Centre National de la Recherche Scientifique, Avenue O. Messiaen, F-72085 Le Mans Cedex 9, France}

\date{\today}

\begin{abstract}
We study the two-dimensional extension of the Su-Schrieffer-Heeger model in its higher order topological insulator phase, which is known to host corner states. Using the separability of the model into a product of one-dimensional Su-Schrieffer-Heeger chains, we analytically describe the eigen-modes, and specifically the zero-energy level, which includes states localized in corners. We then consider networks with disordered hopping coefficients that preserve the chiral (sublattice) symmetry of the model. We show that the corner mode and its localization properties are robust against disorder if the hopping coefficients have a vanishing flux on appropriately defined super plaquettes. We then show how this model with disorder can be realised using an acoustic network of air channels, and confirm the presence and robustness of corner modes. 
\end{abstract}



\maketitle


%
%

\section{Introduction}

Topological insulators have attracted considerable attention in recent years, with a wealth of new topological states of matter that have been discovered~\cite{Hasan10,Budich13,Ryu10,Fu11,Schindler18}. Moreover, these concepts have been applied in photonics or acoustics as powerful tools to control wave propagation~\cite{Zhang18,Ozawa19,Ma19}. The hallmark of topological insulators is the presence of boundary states, with robust propagation properties. Two main classes of topological insulators can be distinguished. In strong topological systems, boundary states are immune to disorder, and hence display robust unidirectional propagation~\cite{Hasan10,Budich13,Bernevig}. On the contrary, in weak topological systems, which rely on translation invariance~\cite{Fu07,Ryu10,Claes20}, it is expected that boundary states will lose their propagation properties upon introducing disorder, for instance through Anderson localization. 

More recently, a new type of topological insulators was introduced: higher order topological insulators~\cite{Slager15,Benalcazar17,Benalcazar17b,Schindler18,Khalaf18}. While a $d$-dimensional topological insulator hosts $d-1$-dimensional boundary states, a $n^{\rm th}$ order topological insulator has $(d-n)$-dimensional boundary states. For instance, two dimensional systems can host topologically protected localized states at their corners, as was observed in kagome~\cite{ElHassan19,Ni19,Xue19} or square lattices~\cite{SerraGarcia18,Imhof18,Mittal19,Qi20,Cerjan20}. However, higher order topological insulators fall into the category of weak topological insulators, and hence, one should expect the topological protection to be broken when adding disorder. 

In this work, we analyze a two-dimensional extension of the well-known Su-Schrieffer-Heeger (SSH) model on a square lattice. This model has been studied in various works~\cite{Liu17,Liu18,Obana19}, in particular it was shown to be a higher order topological insulator hosting localized states at the corners~\cite{Xie18,Ota19,Zhu20,Xu20}, which co-exist with extended bulk ones as bound states imbedded in the continuum~\cite{Chen19,Benalcazar20,Cerjan20}. However, the robustness of these corner modes against disorder has not been thoroughly studied so far, in particular when disorder breaks translation invariance. As a weak topological insulator, and because the corner mode is embedded in the continuum, one would expect disorder to hybridize the corner mode with bulk modes thereby suppressing its localization properties. We point out that this situation contrasts with that of corner modes in quadrupole topological insulators~\cite{Benalcazar17b,SerraGarcia18,Imhof18,Mittal19,Qi20}, where the corner modes lies inside the gap, and is therefore expected to be more robust, as was recently showed in~\cite{Li20,Yang20,Zhang20}.

On the contrary, we show that a corner mode of the 2D SSH model is robust to a large class of disorder. It is robustly localized if it has support on the same sublattice as in the periodic case. This is guaranteed if the disorder satisfies a simple condition: appropriately defined super plaquettes must have a vanishing flux. We then study an acoustic realisation of the 2D SSH model with disorder hosting corner modes, by extending the setup of~\cite{Zheng19} that uses networks of air channels to disordered configurations. Note that most acoustic realizations of higher order topological insulators are based on coupled resonators and rely on a tight binding approximation. This usually restrict the range of validity of the discrete model (such as 2D SSH) to a narrow band of frequency. On the contrary, our approach allows for a broad band correspondence. 

The paper is organized as follows. In section II we present the 2D SSH model without disorder (clean network). We discuss analytic solutions and energy level degeneracy in finite rectangular networks. In section III we study the effect of disorder. We derive a general expression of the corner mode in disorder with vanishing fluxes, and then compare the localization properties of the zero energy mode for several disorder types and strengths. In section IV, we present the acoustic setup and confirm the presence of robust corner modes.

%
%
\section{Model and separability}
\begin{figure*}[t!]
\centering
\begin{minipage}{0.32\textwidth}
\includegraphics[width=\textwidth]{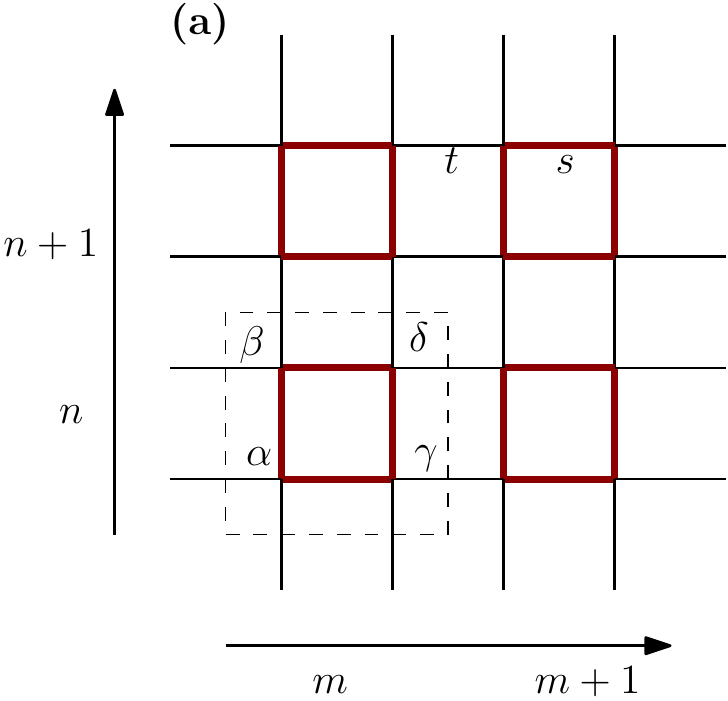}
\end{minipage}
\hspace{5pt}
\begin{minipage}{0.32\textwidth}
\centering
\includegraphics[width=0.5\textwidth]{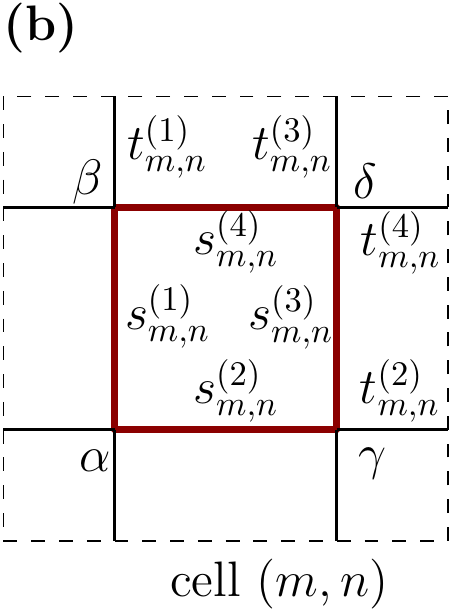}
\end{minipage}
\begin{minipage}{0.32\textwidth}
\includegraphics[width=\textwidth]{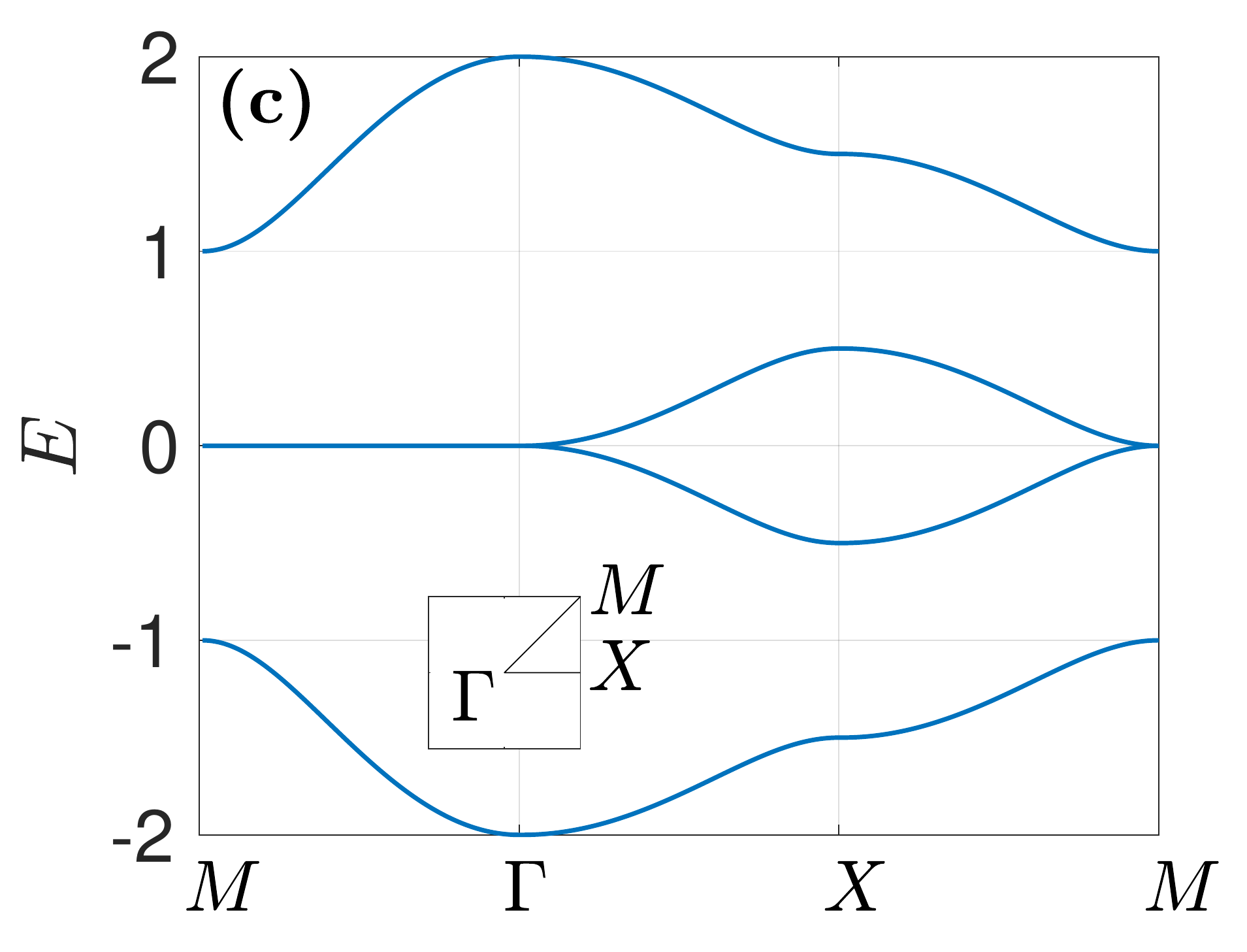}
\end{minipage}
\caption{Schematic representation of the two-dimensional SSH model. (a) General structure. (b) Zoom on a unit cell with indices for unequal hopping coefficients. (c) Dispersion relation.
}
\label{2D_SSH_Fig} 
\end{figure*}
The 2D SSH system is a natural generalisation of the SSH model~\cite{Liu17}, and consists of a square lattice with staggered hopping coefficients (see Fig.~\ref{2D_SSH_Fig}(a)): intracell coefficients $s$ different from intercell ones $t$. A solution $\Phi$ with a given energy $\varep$ solves the eigenvalue problem $H_0 \cdot \Phi = \varep \Phi$ with the Hamiltonian 
\bea
H_0 &=& \sum_{m,n} s |m,n; \alpha \rangle \langle m,n; \beta | + t |m,n; \alpha \rangle \langle m,n-1; \beta | \nonumber \\
&& + s |m,n; \beta \rangle \langle m,n; \delta | + t |m,n; \beta \rangle \langle m-1,n; \delta | \nonumber \\
&& + s |m,n; \delta \rangle \langle m,n; \gam | + t |m,n; \delta \rangle \langle m,n+1; \gam | \nonumber \\
&& + s |m,n; \gam \rangle \langle m,n; \alpha | + t |m,n; \gam \rangle \langle m+1,n; \alpha | \nonumber \\
&& + \mathrm{h.c.}, \label{2D_SSH_cleanH}
\eea
where $(m,n)$ are the lattice indices and $\alpha, \beta, \gamma, \delta$, the intracell indices (see Fig.~\ref{2D_SSH_Fig}(b)). In section~\ref{Acoustic_Sec} we will present a simple acoustic realisation of this model based on a network of air channels. Anticipating this realisation, we impose the following restrictions on the hopping coefficients: they are real with $s>0$, $t>0$, and $t+s=1$. Notice however that negative coefficients can be obtained using coupled resonators~\cite{SerraGarcia18,Qi20}. In the following, a solution will be represented either by the vector $\Phi = \sum_{m,n} \sum_\mu \phi_\mu^{m,n} |m,n;\mu \rangle$ containing all field values, or by a set of 4-vectors $\Phi^{m,n} = (\phi_\alpha^{m,n}, \phi_\beta^{m,n}, \phi_\gamma^{m,n}, \phi_\delta^{m,n})^T$ gathering the field values of all sites within a unit cell and depending on the cell indices $(m,n)$. 

We start by analyzing Bloch wave solutions of an infinite network, $\Phi^{m,n} = \bar \phi e^{i m q_x + i n q_y}$, with $\mathbf{q} = (q_x,q_y)$. The corresponding Bloch Hamiltonian then reads 
\be
h(\mathbf{q}) = \bmat 0 & s + t e^{-iq_y} & s + t e^{-iq_x} & 0 \\ s + t e^{iq_y} & 0 & 0 & s + t e^{-iq_x} \\ s + t e^{iq_x} & 0 & 0 & s + t e^{-iq_y} \\ 0 & s + t e^{iq_x} & s + t e^{iq_y} & 0 \emat . 
\ee
The eigenvalue problem of the Bloch Hamiltonian gives us the dispersion relation of the network, shown in Fig.~\ref{2D_SSH_Fig}(c). In fact, this dispersion relation can be cast under a rather simple form: 
\be
\varep = \pm \left| s + t e^{iq_x} \right| \pm \left| s + t e^{iq_y} \right| , 
\ee
with the four combinations for the $\pm$'s giving us the four branches (see Fig.~\ref{2D_SSH_Fig} (c)). We see that the dispersion relation takes the specific form of a separable system~\cite{Benalcazar20,Zhu20,Cerjan20}, that is $\varep(\mathbf{q}) = \varep_x(q_x) + \varep_y(q_y)$. Moreover, this property extends to the full Hamiltonian of finite (or semi-infinite) networks if the edges of the network are horizontal or vertical, in which case they do not break separability.

\subsection{Finite networks: Eigenmodes}
\label{Finite_Net_Sec}
\begin{figure*}[t!]
\centering
\includegraphics[width=0.9\textwidth]{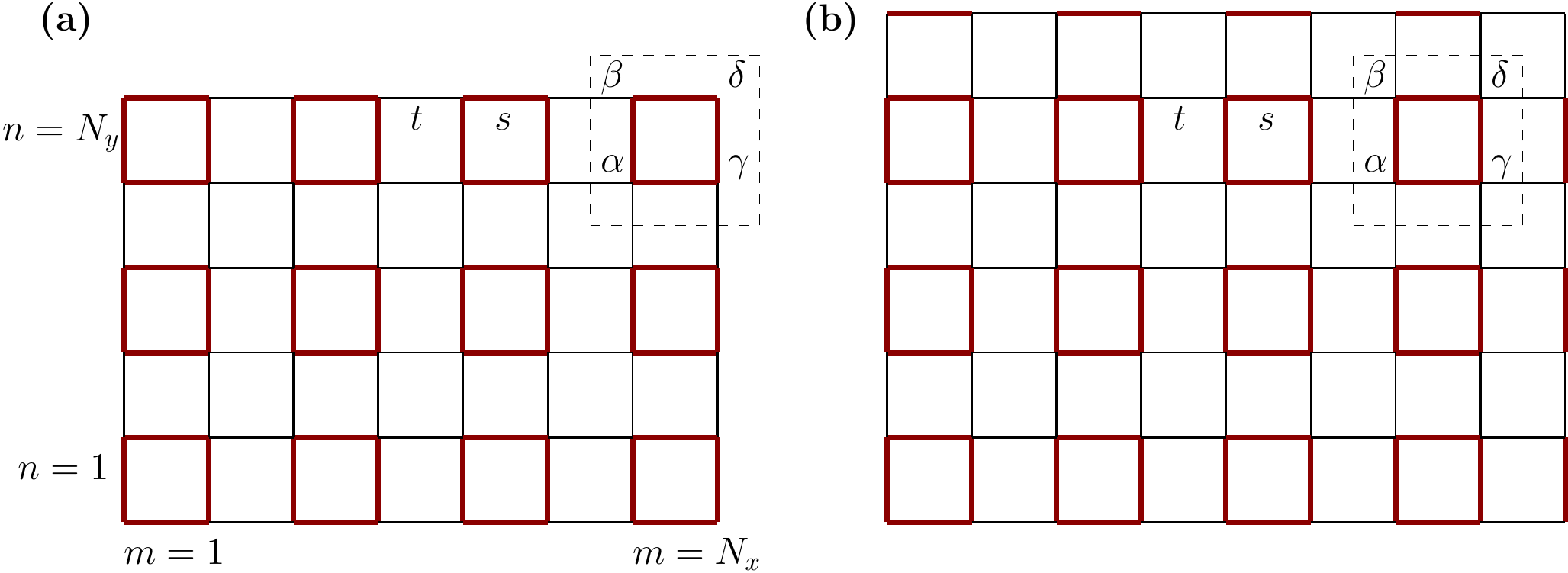}

\includegraphics[width=0.8\textwidth]{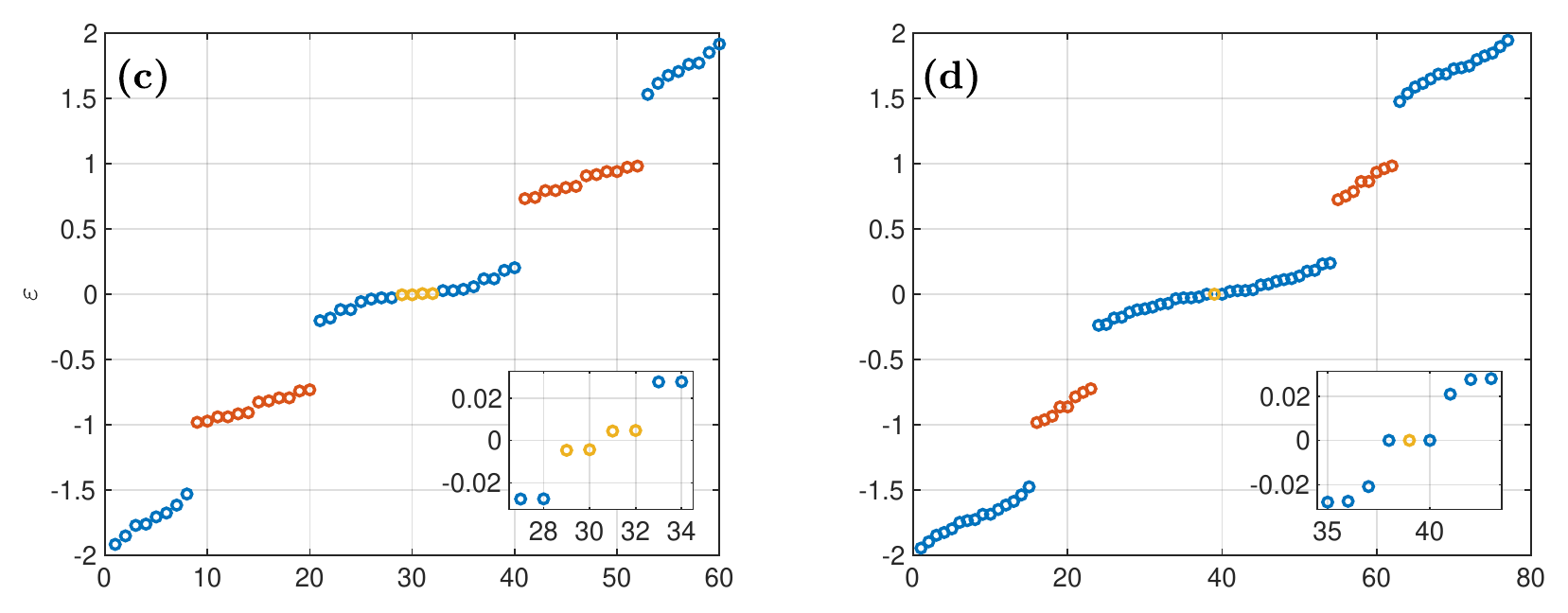}
\caption{(a-b) Representation of finite two-dimensional SSH networks for $N_x=5$ and $N_y=3$. (c-d) Spectrum of the networks showing bulk waves (blue), edge waves (red) and corner modes (yellow) (a,c) Canonical network. (b,d) Asymmetrized network.}
\label{2D_SSH_Finite_Fig} 
\end{figure*}

To study the properties of corner modes, we need to introduce the edges of the sample in our description, and hence we will now consider finite size networks with open boundary conditions. There are basically two ways to cut a rectangular 2D SSH network. The first way is take an integer number of unit cells: $N_x$ horizontally and $N_y$ vertically, as shown in Fig.~\ref{2D_SSH_Finite_Fig}(a). We call this a canonical network. In this case there are two different topological phases: if $s>t$ the network is trivial, without edge waves or corner modes, on the contrary if $s<t$ the network is topological and there are edge waves on the four edges and corner modes in the four corners~\cite{Liu17,Obana19}. An alternative is to add an extra vertical and/or horizontal SSH chain at the edge of the network. As we show in Fig.~\ref{2D_SSH_Finite_Fig}-(b), we will consider a network with an extra chain on the upper edge and the right edge, which we call an asymmetrized network. This amounts to adding an extra site at the end of the corresponding horizontal and vertical SSH chains. Such chains host a unique edge state (see appendix~\ref{1D_App} for details), and as a result, the network of Fig.~\ref{2D_SSH_Finite_Fig}(b) with $s<t$ has edge waves only on the left and lower edges, and a unique corner mode at the lower-left corner. This property of having a unique corner mode is rather convenient to single it out from the rest of the midband, and for that reason, we will mostly investigate this type of network. Nonetheless, our main conclusions remain valid for both types of networks. In the following, unless otherwise specified we will assume $s<t$.

In rectangular finite networks, we can classify all eigenmodes using separability and the knowledge of the 1D SSH chain. Indeed, a complete set of solution can subsequently be obtained by looking at vectors under the form of a tensor product: 
\be \label{Full_2D_TensorProd}
\Phi^{m,n} = \bmat \phi_\alpha^{m,n} \\ \phi_\beta^{m,n} \\ \phi_\gamma^{m,n} \\ \phi_\delta^{m,n} \emat = \bmat \psi_A^m \varphi_A^n \\ \psi_A^m \varphi_B^n \\ \psi_B^m \varphi_A^n \\ \psi_B^m \varphi_B^n \emat \doteq \psi^m \otimes \varphi^n. 
\ee
The vector $\Phi$ is a solution of the 2D SSH model \eqref{2D_SSH_cleanH} if both factors $\psi = \sum_{m} \sum_{\mu} \psi_\mu^{m} |m;\mu \rangle$ and $\varphi = \sum_{n} \sum_{\mu} \varphi_\mu^{n} |n;\mu \rangle$ are solutions of a 1D SSH chain: $\varep_x \psi = H_{0x} \cdot \psi$ and $\varep_y \varphi = H_{0y} \cdot \varphi$, with 
\bea
H_{0x} &=& \sum_{m=1}^{N_x} s |m, B\rangle \langle m, A | + t |m, B\rangle \langle m+1, A | + \mathrm{h.c.} , \nonumber \\ 
H_{0y} &=& \sum_{n=1}^{N_y} s |n, B\rangle \langle n, A | + t |n, B\rangle \langle n+1, A | + \mathrm{h.c.}, \label{xy_Ham}
\eea
and 
\be
\varep = \varep_x + \varep_y. 
\ee
In other words, the 2D Hamiltonian $H_0$ can be written as $H_0 = H_{0x} \otimes I_{2N_y+1} + I_{2N_x+1} \otimes H_{0y}$ with $I_N$ the $N\times N$ identity matrix~\footnote{This is for the configuration of Fig.~\ref{2D_SSH_Finite_Fig}(b). The same is true for Fig.~\ref{2D_SSH_Finite_Fig}(a) with $H = H_{0x} \otimes I_{2N_y} + I_{2N_x} \otimes H_{0y}$, and correspondingly, without the last term $t |N_{x/y}, B\rangle \langle N_{x/y}+1, A |$ in \eq{xy_Ham}.}. Therefore, every rectangular network (possibly infinite in some direction) of 2D SSH can be fully characterized by looking at the two corresponding 1D chains. In appendix~\ref{1D_App}, we recall the main properties of SSH chains. 

Using separability, the topological structure of the 2D SSH model is directly inherited from that of the SSH chain, and in particular its higher order topological insulator character. The product state $\Phi = \psi \otimes \varphi$ belongs to one of three classes: 
\bi
\item If both $\psi$ and $\varphi$ are (1D) bulk waves, then $\Phi$ is a bulk propagating wave with the Bloch wave vector $\mathbf q = q_x \mathbf{e_x} + q_y \mathbf{e_y}$. 
\item If $\psi$ is a bulk wave and $\varphi$ is an edge state ($\varep_y = 0$), then $\Phi$ is an edge wave localized on a horizontal edge. Similarly, if $\psi$ is an edge state and $\varphi$ is bulk wave, then $\Phi$ is an edge wave localized on a vertical edge. 
\item If both $\psi$ and $\varphi$ are edge modes ($\varep_x = 0$ and $\varep_y = 0$), then $\Phi$ is a corner mode. 
\ei
In Fig.~\ref{2D_SSH_Finite_Fig}(c,d), we show the full spectrum for both configurations: canonical and asymmetrized networks respectively. 

\subsection{Finite networks: level degeneracy}
\label{Finite_Dege_Sec}
We now discuss the degeneracy of energy levels, and in particular for edge waves and the zero energy level. We focus on asymmetrized network (Fig.~\ref{2D_SSH_Finite_Fig}(b)), where this can be done explicitly. The main ingredient is that a one-dimensional asymmetrized SSH chain with $N$ cells has energy levels given by a simple expression: 
\be \label{1D_EnergyLevels}
\varep_j = \pm \left|s + t \exp\left(i \frac{j \pi}{N+1} \right) \right|, 
\ee
with $j=1..N$, plus a unique zero-mode $\varep_0 = 0$ (see appendix~\ref{1D_App}). Using separability, the 2D network has energy levels of the form $\varep = \varep_{j_x} + \varep_{j_y}$ (bulk waves), $\varep=\varep_{j_x}+0$ or $\varep=0+\varep_{j_y}$ (edge waves), and $\varep=0+0$ (corner mode), with $j_x=1..N_x$ and $j_y=1..N_y$. 

Let us start by discussing edge waves. Using \eq{1D_EnergyLevels}, we see that degenerate energies for edge waves can only happen if $\varep_{j_x} = \varep_{j_y}$ for some $j_x$ and $j_y$, which corresponds to a left edge wave having the same energy as a down edge wave. If $N_x = N_y$, this is satisfied by swapping the roles of $x$ and $y$ to obtain the same eigenvalue meaning in that case every edge eigenvalue is doubly degenerate. If $N_x \neq N_y$, we see that $\varep_{j_x} = \varep_{j_y}$ only if $e^{ij_x \pi/(N_x+1)} = e^{ij_y \pi/(N_y+1)}$ (one can use \eq{1D_SSH_DispRel}). Hence, we must find $j_x \in \{1,2..N_x\}$ and $j_y \in \{1,2..N_y\}$ such that 
\be
j_x (N_y+1) = j_y (N_x+1). 
\ee
We now introduce the greatest common divisor $N_d = \mathrm{gcd}(N_x+1, N_y+1)$ so that $N_x+1 = N_d n_x$ and $N_y+1 = N_d n_y$, with $n_x$ and $n_y$ co-prime. The above equality becomes $j_x n_y = j_y n_x$, and hence we have the following pairs of solutions $(j_x,j_y)$: 
\bsub \bea
&(n_x, n_y),&  \\
&(2n_x, 2n_y),& \\
&\vdots & \nonumber \\
&((N_d-1)n_x, (N_d-1)n_y),&  
\eea \esub
and one cannot go further since one would have $j_x = N_d n_x = N_x +1 > N_x$. We then conclude that we have $N_d-1$ pairs of doubly degenerate edge modes of positive energy and $N_d-1$ pairs of doubly degenerate edge modes of negative energy. 

With a similar line of thought, we can obtain the degeneracy of the zero energy level. There is always at least one zero mode: the corner mode of \eq{Clean_CornerMode}, corresponding to $\varep=0+0$. But we can also have bulk waves with zero energy $\varep = \varep_{j_x} + \varep_{j_y}$, if $\varep_x = -\varep_y$. Using chiral symmetry of the spectrum, this leads to the same condition as above. Hence, one can directly conclude that there are $2(N_d-1)$ bulk modes of zero energy ($N_d-1$ with $\varep_x>0$ and $N_d-1$ with $\varep_x<0$). Including the corner state, this leaves us with $2N_d - 1$ zero energy modes. This result is well illustrated in Fig.~\ref{2D_SSH_Finite_Fig}(c,d). For the canonical network there is no degeneracy, as we see in Fig.~\ref{2D_SSH_Finite_Fig}(c). For the asymmetrized network, $N_d=2$, and we observe three modes at zero energy: the corner mode and two bulk waves, as shown in Fig.~\ref{2D_SSH_Finite_Fig}(d).

\subsection{Finite networks: corner modes}

\begin{figure*}[t!]
\centering
\includegraphics[width=\textwidth]{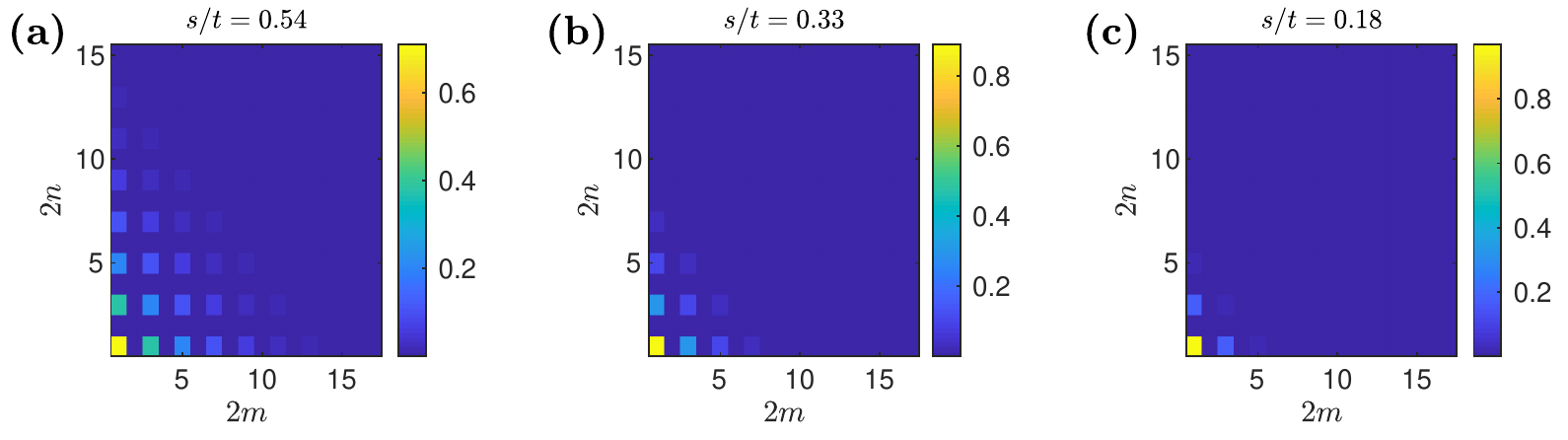}
\caption{(a-c) Modulus of the components of the corner mode for an asymmetrized network (Fig.~\ref{2D_SSH_Finite_Fig}(b,d)) and different values of the hopping coefficients. (a) $s=0.35$ and $t=0.65$. (b) $s=0.25$ and $t=0.75$. (c) $s=0.15$ and $t=0.85$. 
}
\label{2D_Periodic_Modes_Fig} 
\end{figure*}

A key aspect of the SSH model and its 2D generalization is that it is chiral symmetric because it consists of two sublattices with hopping only between each other. One sublattice is made of the $(\alpha, \delta)$ sites and the other of $(\beta, \gamma)$ sites. This bipartite structure leads to a chiral symmetry operator $\Gam$ such that $\Gam \cdot H_0 + H_0 \cdot \Gam = 0$ (see appendix~\ref{Chiral_App}). The main consequences of chiral symmetry are twofold. First, eigenvectors of the Hamiltonian come in pairs of chiral partners with opposite energies. Second, zero energy modes vanish on one of the two sublattices. Remarkably, corner modes have vanishing amplitudes on three sites per unit cell, while chirality imposes only two amplitudes to vanish (as we explain in appendix~\ref{Chiral_App}, this can be seen as the result of horizontal and vertical partial chiral symmetries). For an asymmetrized network (Fig.~\ref{2D_SSH_Finite_Fig}(b)) there is a unique corner mode (noted $\Phi_0$), which has an explicit expression as a product state, as in \eq{Full_2D_TensorProd}: 
\be \label{Clean_CornerMode}
\Phi_0^{m,n} = A \bmat 1 \\ 0 \\ 0 \\ 0 \emat \left(-\frac{s}{t}\right)^{m+n} , 
\ee
where $A$ is a normalization constant, fixed by requiring $|| \Phi_0 || = 1$. Equation~\eqref{Clean_CornerMode} shows that the corner mode has support only on the $\alpha$-sites. As we shall see when introducing disorder, this specific sublattice structure of the corner mode is key to its robustness. Moreover, in an asymmetrized network \eq{Clean_CornerMode} automatically satisfies the open boundary conditions~\footnote{This can be seen by adding nearest neighbour ghost sites around the network, such that the boundary condition is equivalent to the field amplitudes vanishing on these ghost sites. For an asymmetrized network as in Fig.~\ref{2D_SSH_Finite_Fig}(b), these ghost sites are all $\beta$, $\gam$ or $\delta$ and hence the constructed corner state of \eq{Clean_CornerMode} vanishes on them. \label{BC_ftn}}. 

This corner mode is shown in Fig.~\ref{2D_Periodic_Modes_Fig} for different values of the hopping coefficients. When $s<t$, the upper and right boundaries can be send to infinity and \eq{Clean_CornerMode} gives a localized (exact) solution, i.e. a bound state in the continuum. From this result we also conclude that on a canonical network (Fig.~\ref{2D_SSH_Finite_Fig}(a)) with $s<t$ (topological phase), \eq{Clean_CornerMode} is an approximate solution, with three other solutions of similar form in the three other corners. Finite size effects lift the degeneracy due to evanescent coupling, and the eigenmodes are then given by linear combinations of the four corner modes with appropriate symmetries, as studied e.g. in~\cite{Zhu20}. However, because the evanescent coupling is exponentially suppressed for large networks, corner modes for different corners can be treated independently. Therefore, we expect that all our results obtained in asymmetrized networks will hold for canonical networks, modulo appropriate changes of the role of intracell indices.

%
%
\section{Effect of disorder on corner modes}
The 2D SSH model has been shown in several works~\cite{Xie18,Chen19} to be a higher order topological insulator. However, the robustness of the corner mode is particularly non-trivial since its energy lies inside the middle band, at $\varep = 0$. Therefore, upon introducing disorder, we could expect the corner mode to hybridize with the bulk waves of the mid-band, and lose its localization property as soon as separability is broken~\cite{Benalcazar20,Cerjan20}. As we shall see, this is not the case, and the corner mode stays robust under a much milder condition: it must have support on the same sublattice as in the clean case.

Our analysis is focused on the properties of the corner mode when disorder is added on the hopping coefficients. This type of disorder, also referred to as off-diagonal disorder, does not break the chiral symmetry of the network, in contrast for example to on-site energy disorder. In several works~\cite{Chen19,Benalcazar20,Cerjan20,Li20,Yang20,Zhang20}, it was shown that chiral symmetry is a necessary condition to have robust corner modes, which is why we focus on chiral preserving disorders. 

We consider a finite asymmetrized network as in Fig.~\ref{2D_SSH_Finite_Fig}(b). An advantage of such a network is that there is always at least one zero-energy state even in the presence of disorder. This can be seen by noticing that the first sublattice $(\alpha, \delta)$ contains one additional site with respect to the second sublattice $(\beta,\gam)$, and a general property of chiral systems is that if one sublattice contains more sites than the other, there are as many zero energy solutions that vanish on the minority sublattice than the difference in the number of sites~\cite{Inui94}. Therefore, upon introducing disorder chirality guarantees that at least one zero energy state is present in asymmetrized networks. 

In a clean network, one such zero-energy solution is given by the corner mode of \eq{Clean_CornerMode} as a product of two one-dimensional SSH edge states. However, in square shaped networks ($N_x = N_y$), the zero energy level is highly degenerate, as we saw in section~\ref{Finite_Dege_Sec}. To single out the corner mode, several strategies have been proposed, such as introducing diagonal disorder~\cite{Chen19} or dissipation in the bulk~\cite{Benalcazar20}. A simpler alternative is to break the symmetry between $x$ and $y$ by considering rectangular networks. From now on, we will assume $N_x = N_y+1$, in which case we know from section~\ref{Finite_Dege_Sec} that there is always a unique zero energy solution. To investigate the robustness of the localization properties of the corner mode against disorder, we follow the unique zero energy mode and identify under what conditions it is well localized in the corner. 

Using the notations of Fig.~\ref{2D_SSH_Fig}(b), the disordered Hamiltonian reads  
\bea
H &=& \sum_{m,n} |m,n; \alpha \rangle \left(s_{m,n}^{(1)} \langle m,n; \beta | + t_{m,n-1}^{(1)} \langle m,n-1; \beta | \right) \nonumber \\
&& + |m,n; \beta \rangle  \left( s_{m,n}^{(4)} \langle m,n; \delta | + t_{m-1,n}^{(4)} \langle m-1,n; \delta | \right) \nonumber \\
&& + |m,n; \delta \rangle \left( s_{m,n}^{(3)} \langle m,n; \gam | + t_{m,n}^{(3)} \langle m,n+1; \gam | \right) \nonumber \\
&& + |m,n; \gam \rangle \left( s_{m,n}^{(2)} \langle m,n; \alpha | + t_{m,n}^{(2)} \langle m+1,n; \alpha | \right) \nonumber \\
&& + \mathrm{h.c.} \label{2D_SSH_Disorder}
\eea
In the following, we will consider and compare three types of disorder: general unconstrained disorder, separable disorder, and zero flux disorder. 

\begin{figure}[htp]
\centering
\includegraphics[width=0.8\columnwidth]{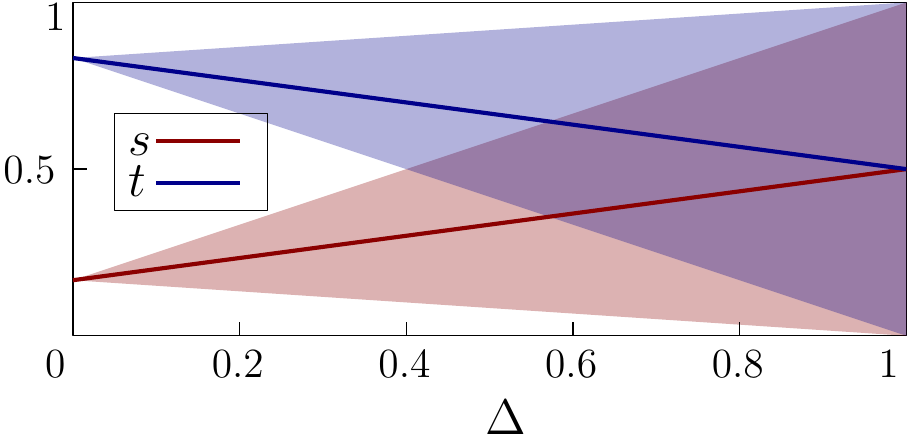}
\caption{Probability distribution of the hopping coefficients for increasing disorder strength $\Delta$. The shades indicate the range of the uniform distributions, and the solid lines show the mean values. 
}
\label{Hopping_Distrib_Fig} 
\end{figure}

To discuss this, we build a disordered Hamiltonian $W$, with coefficients $s$ and $t$ randomly and independently picked between 0 and 1. We then look at interpolated Hamiltonians between the clean Hamiltonian $H_0$ of \eq{2D_SSH_cleanH} and the fully disordered one $W$. Here, we investigate a family of disordered Hamiltonians with unconstrained disorder of the form: 
\be \label{2D_Disorder_H}
H(\Delta) = (1-\Delta) H_0 + \Delta W, 
\ee
with $\Delta \in [0,1]$. The constructed Hamiltonian is of the form of \eq{2D_SSH_Disorder} with random hopping coefficients uniformly distributed over an interval of size $\Delta$ and of mean values $\langle s \rangle = (1-\Delta) s_0 + \Delta/2$ and $\langle t \rangle = (1-\Delta) t_0 + \Delta/2$, as represented in Fig.~\ref{Hopping_Distrib_Fig}. Doing so, the hopping coefficients of $H$ stay between 0 and 1, in order to be compatible with the acoustic realisation to be presented below. Notice also that the constructed disorder preserves chiral symmetry, that is for any $\Delta$ we have $\Gam \cdot H + H \cdot \Gam = 0$. Equation~\eqref{2D_Disorder_H} gives us the Hamiltonian in an unconstrained disorder. As detailed in appendix~\ref{Disorder_App}, the two other disorder types are built similarly. 

%
%
\subsection{Unconstrained disorder}

\begin{figure*}[t!]
\centering
\includegraphics[width=0.9\textwidth]{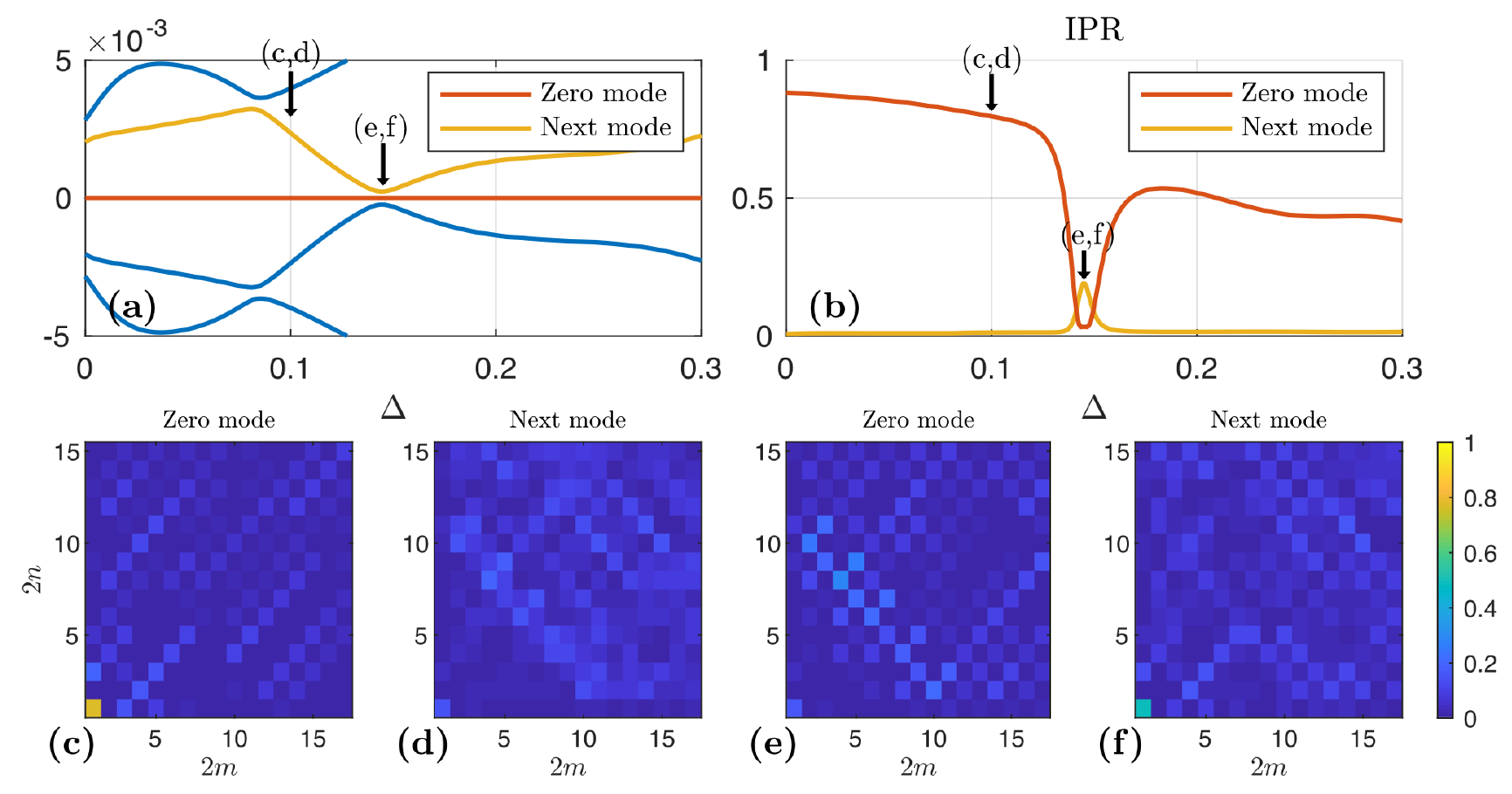}
\caption{Finite network with $N_x = 8$, $N_y = 7$, $s_0 = 0.15$ and $t_0 =0.85$. (a) Spectrum of $H(\Delta)$ with a given disorder realisation of $W$ and varying the disorder level $\Delta$. (b) IPR of the zero-mode (red line) and the next mode (yellow line), which refers to the mode with lowest positive nonzero energy. (c-f) Modulus of the components of the zero-mode (c and e) and the next mode (d and f) for two values of disorder: $\Delta=0.1$ (c-d) and $\Delta=0.145$ (e-f). The two values are shown by an arrow in (b). 
}
\label{CornerMode_SmallNet_Fig} 
\end{figure*}

We first want to understand how the corner mode interacts with the midband in a general disorder. Chiral symmetry guarantees that its frequency is robust since as we saw earlier there is a zero-mode for any disorder. However, when turning disorder on, this zero-mode might lose its localization properties by hybridising with bulk waves. To characterize how localized the zero-mode is, we introduce the Inverse Participation Ratio (IPR)~\cite{Thouless74}. The IPR of a mode $\Phi$ is defined as 
\be
I(\Phi) = \sum_{m,n} \sum_{\mu = \alpha, \beta, \gam, \delta} |\phi_\mu^{m,n}|^4, 
\ee
where the mode must be normalized, i.e. $\sum |\phi_\mu^{m,n}|^2 = 1$. It is easy to see that the IPR is always between 0 and 1. When the mode is spread in the bulk, it has a low IPR, and on the contrary, if the mode has a few nonzero components, its IPR is higher. We will also consider a variant of the IPR, where we add a weight to penalise sites far from the lower-left corner. Explicitly, we define 
\be
I_c(\Phi) = \sum_{m,n} \sum_{\mu = \alpha, \beta, \gam, \delta} \frac{|2 \phi_\mu^{m,n}|^4}{(m+n)^4}, 
\ee
where the factor $2$ is here so that the weight is unity for the most lower-left corner $(m,n)=(1,1)$. This weighted IPR will allow us to discriminate whether the mode is localized near the corner or in bulk (due to disorder), in which case $I_c$ starts to be lower than $I$. 

We now analyze the change of IPR of the zero-mode in disordered networks when continuously increasing the strength of disorder. The results are shown in Fig.~\ref{CornerMode_SmallNet_Fig}. Because we consider a reasonably small network, the energy spacing near $\varep=0$ is still appreciable and hence, the zero-mode interacts essentially with the first pair of modes with nonzero energy. When increasing the disorder strength $\Delta$, the eigenvalues move and repulse each other. This is the usual avoided crossing phenomenon, Fig.~\ref{CornerMode_SmallNet_Fig}(a). This applies in particular to the zero-mode (red in Fig.~\ref{CornerMode_SmallNet_Fig}(a)) and the pair of modes with the smallest nonzero energies (we call the one with positive $\varep$ ``next mode'' and show it in yellow in Fig.~\ref{CornerMode_SmallNet_Fig}(a)). What is remarkable is that this avoided crossing in accompanied by a sudden drop of the IPR, as shown in Fig.~\ref{CornerMode_SmallNet_Fig}(b). Far from the avoided crossing, the zero-mode is localized in the corner and the next mode is a bulk wave, as in Fig.~\ref{CornerMode_SmallNet_Fig}(c-d). Near the crossing, the two modes are swapped: the zero-mode spreads in the bulk while the next mode is localized in the corner, as in Fig.~\ref{CornerMode_SmallNet_Fig}(e-f). By interacting with midband modes, the corner mode acquired a nonzero energy, despite the fact that the disorder preserves chiral symmetry. In fact, by having a nonzero energy, chiral invariance implies that there is now a pair of corner modes with opposite energy (chiral partners) but with nonzero components on the full lattice, and not only the $(\alpha,\delta)$-sublattice. This explains why the corresponding IPR is not as high as that of the zero-mode far from the crossing. When the size of the system is larger, the zero mode interacts with more bulk waves, and hence loses its localization properties more rapidly with increasing disorder.

%
%
\subsection{Separable disorder}
In general, disorder breaks the separability of the Hamiltonian, and eigenvectors can no longer be found as product states as in \eq{Full_2D_TensorProd}. However, there is a particular disorder structure that maintains the decomposition and hence allows for a simple construction of solutions from that of 1D chains. To see this, we reverse the logic and consider two disordered 1D chains: 
\be
H_{x} = \sum_{m=1}^{N_x} s_m^{\rm x} |m, B\rangle \langle m, A | + t_m^{\rm x} |m, B\rangle \langle m+1, A | + \mathrm{h.c.} ,  
\ee
and 
\be
H_{y} = \sum_{n=1}^{N_y} s_n^{\rm y} |n, B\rangle \langle n, A | + t_n^{\rm y} |n, B\rangle \langle n+1, A | + \mathrm{h.c.} 
\ee
Now, the tensor product $H_{x} \otimes I_{2N_y+1} + I_{2N_x+1} \otimes H_{y}$ gives the 2D Hamiltonian of \eq{2D_SSH_Disorder} if the hopping coefficients are of the form 
\bsub \label{SeparableDis_eq} \bea
s_{m,n}^{(1)} = s^{\rm y}_n \qquad &\textrm{and}& \qquad t_{m,n}^{(1)} = t^{\rm y}_n, \\
s_{m,n}^{(2)} = s^{\rm x}_m \qquad &\textrm{and}& \qquad t_{m,n}^{(2)} = t^{\rm x}_m, \\
s_{m,n}^{(3)} = s^{\rm y}_n \qquad &\textrm{and}& \qquad t_{m,n}^{(3)} = t^{\rm y}_n, \\
s_{m,n}^{(4)} = s^{\rm x}_m \qquad &\textrm{and}& \qquad t_{m,n}^{(4)} = t^{\rm x}_m , 
\eea \esub
which we refer to as ``separable disorder''. The above form can be stated in simple geometric terms: hopping coefficients corresponding to horizontal (resp. vertical) links must only depend on the horizontal coordinate $m$ (resp. vertical coordinate $n$). 

In this type of disorder, the zero-mode is still given by a product state similar to \eq{Clean_CornerMode}: 
\be \label{Separable_CornerMode}
\Phi_0^{m,n} = A \bmat 1 \\ 0 \\ 0 \\ 0 \emat \prod_{j=1}^m \left(-\frac{s_{j}^{\rm x}}{t_{j}^{\rm x}}\right) \prod_{j'=1}^n \left(-\frac{s_{j'}^{\rm y}}{t_{j'}^{\rm y}}\right) , 
\ee
where $A$ is a normalization constant. Therefore, its localization properties are directly inherited from that of edge states in the corresponding horizontal and vertical SSH chains. Localization at zero energy in chiral one-dimensional lattices is relatively well studied~\cite{Eggarter78,Inui94,Mondragon14}, and many aspects can be understood from the fact that $\ln |\psi^m|$ is a biased random walk. For our purpose, we underline the two main consequences for the corner mode of \eq{Separable_CornerMode} written as a product state. First, as in the clean case it vanishes on all $\beta$, $\gamma$ and $\delta$ sites. Second, the field amplitude decreases for increasing $m$ (resp. $n$) if $\langle \ln|s_{j}^{\rm x}/t_{j}^{\rm x}| \rangle < 0$ (resp. $\langle \ln|s_{j}^{\rm y}/t_{j}^{\rm y}| \rangle < 0$) as in the one-dimensional case~\cite{Mondragon14}. Note that the average is taken over the different cells, not the disorder realization, but the two become equivalent in the limit of large networks.

%
%
\subsection{Zero flux disorder}
\label{AnalyticCorner_Sec}

\begin{figure*}[t!]
\centering
\includegraphics[width=0.8\textwidth]{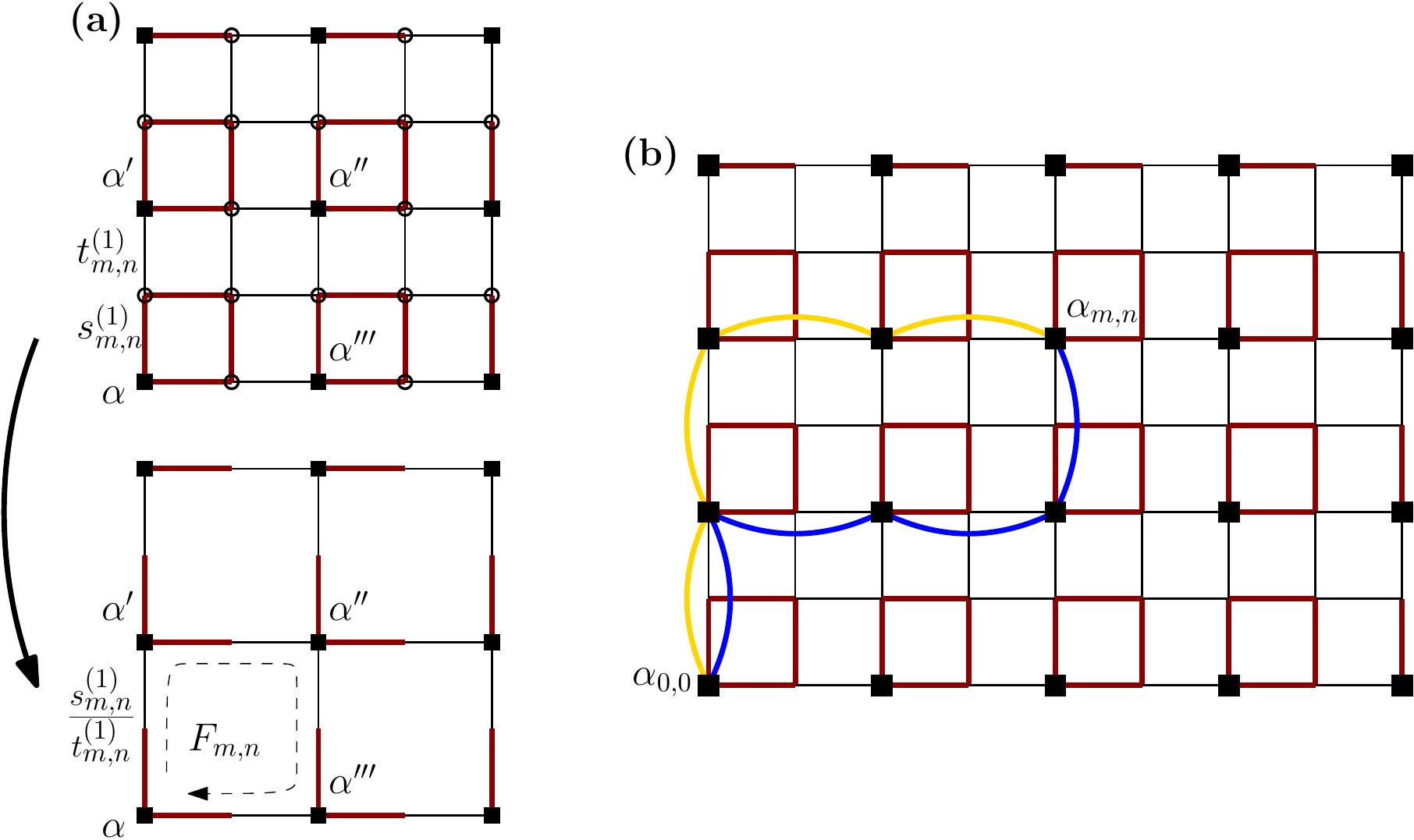}
\caption{(a) Construction of super-plaquettes and the associated flux. A super-plaquette is obtained by taking four neighbouring $\alpha$-sites and deleting links connecting $(\beta,\gam,\delta)$ with each other. $\alpha$ sites (where the field amplitude is non-zero) are marked by black squares and $\beta,\gam$, and $\delta$ (field amplitude is zero) by circles. (b) Representation of two paths (blue and yellow lines) from $\alpha_{0,0}$ to $\alpha_{m,n}$ through neighbouring $\alpha$ sites (marked by black squares). (a-b) Note that the hopping coefficients are disordered, and the red and black lines indicate the one labelled by $s$ or $t$. 
}
\label{2D_CornerPaths_Fig} 
\end{figure*}

The strong robustness of the corner mode against disorder extends far beyond separability, and find its origin in its peculiar sublattice structure, namely having support only on the $\alpha$ sites. To see this, we look for a solution that vanishes on all $(\beta, \gam, \delta)$ sites. Projecting $H$ of \eq{2D_SSH_Disorder} with $\langle m,n; \alpha|$, we see that we must have $\varep=0$, and hence we are left with only two nontrivial equations (see appendix~\ref{CornerMode_App} for more details): 
\bsub \label{2D_DisorderSSH_WithCmode} \bea
0 &=& s_{m,n}^{(1)} \phi_\alpha^{m,n} + t_{m,n}^{(1)} \phi_\alpha^{m,n+1} , \label{2D_DisorderSSH_WithCmode_y} \\
0 &=& s_{m,n}^{(2)} \phi_\alpha^{m,n} + t_{m,n}^{(2)} \phi_\alpha^{m+1,n} . \label{2D_DisorderSSH_WithCmode_x}
\eea \esub
To ease the discussion, we refer to the $\alpha$ site of the cell $(m,n)$ as $\alpha_{m,n}$ (see Fig.~\ref{2D_CornerPaths_Fig}(b)). Equation~\eqref{2D_DisorderSSH_WithCmode_y} gives us the field amplitude on $\alpha_{m,n+1}$ if we know the amplitude on $\alpha_{m,n}$. Similarly, \eq{2D_DisorderSSH_WithCmode_x} allows us to go from $\alpha_{m,n} \to \alpha_{m+1,n}$. Using the two equations we can now relate the amplitude on $\alpha^{m+1,n+1}$ to the one on $\alpha^{m,n}$ in two different ways: either going from $\alpha_{m,n} \to \alpha_{m+1,n} \to \alpha_{m+1,n+1}$ or from $\alpha_{m,n} \to \alpha_{m,n+1} \to \alpha_{m+1,n+1}$. Having a nontrivial solution of equations~\eqref{2D_DisorderSSH_WithCmode} requires the two obtained relations to be compatible, which is true if the hopping coefficients satisfy 
\be \label{ZeroFlux_eq}
F_{m,n} = \ln\left(\frac{s_{m,n}^{(1)}}{t_{m,n}^{(1)}} \frac{s_{m,n+1}^{(2)}}{t_{m,n+1}^{(2)}} \frac{t_{m+1,n}^{(1)}}{s_{m+1,n}^{(1)}} \frac{t_{m,n}^{(2)}}{s_{m,n}^{(2)}} \right) = 0. 
\ee
$F_{m,n}$ represents the flux around a super-plaquette made of neighbouring $\alpha$-sites, as illustrated in Fig.~\ref{2D_CornerPaths_Fig}(a). As we showed, to have a a corner mode with the same sublattice structure as in the clean case \eqref{Clean_CornerMode}, it is necessary that all the fluxes $F_{m,n}$ be trivial. It turns out that this condition of zero fluxes is also sufficient to have a solution with support on $\alpha$-sites only. Indeed, assuming \eq{ZeroFlux_eq} holds, we build the product state 
\be \label{General_CornerMode}
\Phi_0^{m,n} = A \bmat 1 \\ 0 \\ 0 \\ 0 \emat \prod_{\mathcal C(\alpha_{0,0} \to \alpha_{m,n})} \left(-\frac{s^{(\nu)}_{m',n'}}{t^{(\nu)}_{m',n'}}\right) . 
\ee
In this equation, $\mathcal C(\alpha_{0,0} \to \alpha_{m,n})$ is a path in the network going from $\alpha_{0,0}$ to $\alpha_{m,n}$ through neighbouring $\alpha$ sites, as shown in Fig.~\ref{2D_CornerPaths_Fig}(b). Each step from a site $\alpha_{m',n'}$ to a neighbour $\alpha_{m'+1,n'}$ or $\alpha_{m',n'+1}$ is associated with the corresponding ratio of hopping coefficients $s^{(\nu)}_{m',n'}/t^{(\nu)}_{m',n'}$ with $\nu=1,2$ that enters in the product of \eq{General_CornerMode}. Now, the zero flux condition of \eq{ZeroFlux_eq} tells us that the result is independent of the chosen path. This construction is illustrated in Fig.~\ref{2D_CornerPaths_Fig}. It is then straightforward to see that it provides a zero-energy solution of the 2D SSH equations~\eqref{2D_SSH_Disorder}, reducing to equations~\eqref{2D_DisorderSSH_WithCmode}, by choosing an appropriate path: ending by $\alpha_{m,n} \to \alpha_{m,n+1}$ for \eq{2D_DisorderSSH_WithCmode_y} and by $\alpha_{m,n} \to \alpha_{m+1,n}$ for \eq{2D_DisorderSSH_WithCmode_x}. 

Just like in the clean case, \eq{General_CornerMode} always satisfies the boundary conditions in an asymmetrized network (while it only gives an approximate solution for a canonical network in the topological phase, see footnote~\ref{BC_ftn}). However, the localization of the mode of \eq{General_CornerMode} depends on the average behavior of $\ln |s/t|$. If $\langle \ln |s/t|\rangle$ is negative, the amplitude of \eq{General_CornerMode} decreases on average for increasing $(m,n)$, and hence, the mode is localized on the lower left corner. On the contrary, if $\langle \ln |s/t|\rangle$ is positive, \eq{General_CornerMode} is localized on the upper right corner. However, for $\langle \ln |s/t| \rangle < 0$ but large disorder strengths, the lower left values of $\ln |s/t|$ might be positive in some realisations. In that case the amplitude of \eq{General_CornerMode} first increases before decreasing, and the zero-mode leaks further away from the corner. In the extreme case of $\langle \ln |s/t| \rangle = 0$ with disorder, the zero-mode of \eq{General_CornerMode} is anomalously localized, decreasing like $O(e^{-\lam \sqrt m - \lam' \sqrt n})$~\cite{Inui94}, but not specifically in the corner. In appendix~\ref{2DRandomWalk_App}, we explain how this anomalous localization can be understood from random walks, as in the one-dimensional case.

The corner mode construction developed above turns out to be rather general. For instance, the same zero flux condition in similarly constructed super plaquette has been found in~\cite{Poli17} as a condition to preserve a corner mode in the presence of disorder. Furthermore, our construction also applies to other lattice configurations, such as kagome lattices that display corner modes~\cite{Ni19,Xue19} (this is outlined in appendix~\ref{Kagome_App}).

%
%
\subsection{Comparing the three types of disorder}
\label{Num_Disorder_Sec}

\begin{figure*}[t!]
\centering
\includegraphics[width=\textwidth]{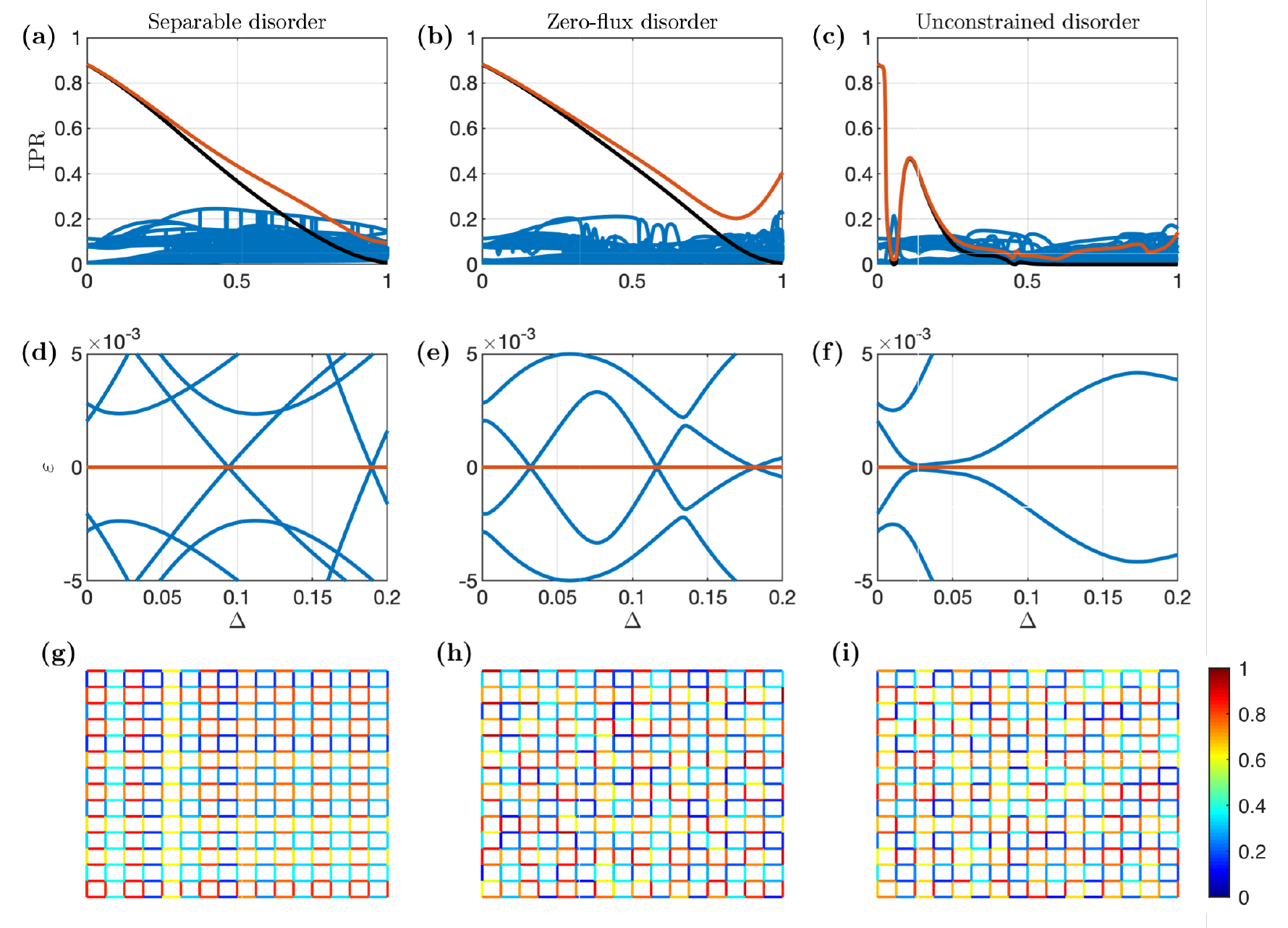}
\caption{Spectrum and IPR for different types of disorder in finite networks with $N_x = 8$, $N_y = 7$, $s_0 = 0.15$ and $t_0 = 0.85$. (a-c) IPR of the zero-mode ($I$ in red and $I_c$ in black) and all the other modes (blue). (d-f) Eigenvalues near 0 as a function of the disorder strength. (g-i) Example of networks for all types of disorder, with color scale showing the hopping coefficient values. We took $\Delta=0.3$. (a,d,g) Separable disorder. (b,e,h) Zero flux disorder defined by \eq{ZeroFlux_eq}. (c,f,i) Unconstrained disorder. 
}
\label{CornerMode_BigNet_Fig} 
\end{figure*}

We now compare three types of disorder: separable, with vanishing fluxes $F_{m,n}$ (\eq{ZeroFlux_eq}), and unconstrained (but still chiral). In Fig.~\ref{CornerMode_BigNet_Fig}(a-c) we show the IPR of the zero-mode (in red) and compare it to the IPR of all other modes (in blue). For all disorder types, the IPR of all modes but the zero-mode tend to increase with disorder strength $\Delta$, due to wave localization. Moreover, in an unconstrained disorder we see that the zero-mode IPR quickly drops until it becomes comparable to all other modes, meaning that it is no longer localized as a corner mode but rather due to wave localization~\cite{Eggarter78,Inui94,Evangelou03}. On the contrary, for separable disorders and disorders obeying the constraint of \eq{ZeroFlux_eq}, the zero-mode stays well localized until high values of the disorder strength. This is a manifestation of the robustness of the corner mode against this types of disorder. At higher disorder strengths, the mode is still localized but around a point that can move away from the corner, as explained in section~\ref{AnalyticCorner_Sec}. This is confirmed by the decrease of $I_c$ compared to $I$ in Fig.~\ref{CornerMode_BigNet_Fig}(a-c). In Fig.~\ref{CornerMode_BigNet_Fig}(d-f), we show the evolution of eigenvalues near zero with the disorder strength $\Delta$. When the disorder is separable, eigenvalues cross without interacting. This is because the corresponding modes do not interact due to the conservation of the transverse wavenumber. When the disorder is not separable but satisfies \eq{ZeroFlux_eq}, eigenvalues repulse each other as in a general disorder, but crossing can occur at $\varep=0$. This suggests that bulk waves see a general disorder, but no longer interact with the corner mode. We also point out that the results shown in Fig.~\ref{CornerMode_BigNet_Fig} are obtained for a single disorder realisation per disorder type, and no average has been performed. Although the exact curves vary for different realisations, the distinct behaviours between the three types of disorder remain. 

\begin{figure*}[t!]
\centering
\includegraphics[width=\textwidth]{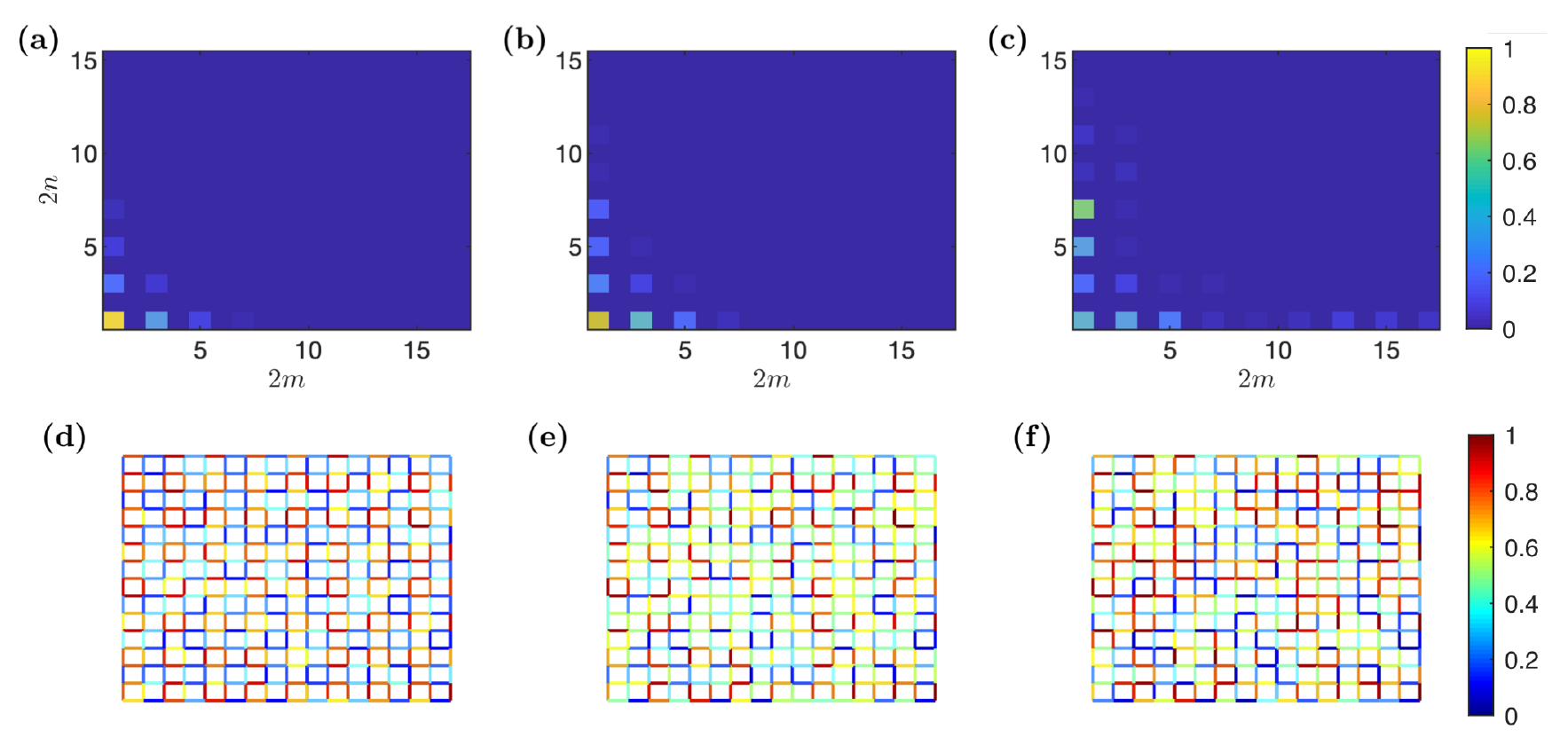}
\caption{(a-c) Modulus of the components of the zero-mode for a disordered network with zero fluxes, with $N_x = 8$, $N_y = 7$, $s_0 = 0.15$ and $t_0=0.85$. (d-f) Representation of the corresponding networks with color scale showing the hopping coefficient values. (a,d) $\Delta = 0.3$. (b,e) $\Delta = 0.6$. (c,f) $\Delta = 0.9$. 
}
\label{CornerMode_draw_Fig} 
\end{figure*}

To further emphasize the robustness in zero flux disorders, we show the zero-mode in Fig.~\ref{CornerMode_draw_Fig}(a-c), with the corresponding disordered network shown in Fig.~\ref{CornerMode_draw_Fig}(d-f). We see that it is well localized in the corner even at high disorder intensities, although the site of maximum amplitude is away from the corner at very high disorder strengths (see Fig.~\ref{CornerMode_draw_Fig}(c)).

%
%
\subsection{Topological defect modes}
\begin{figure*}[t!]
\centering
\includegraphics[width=\textwidth]{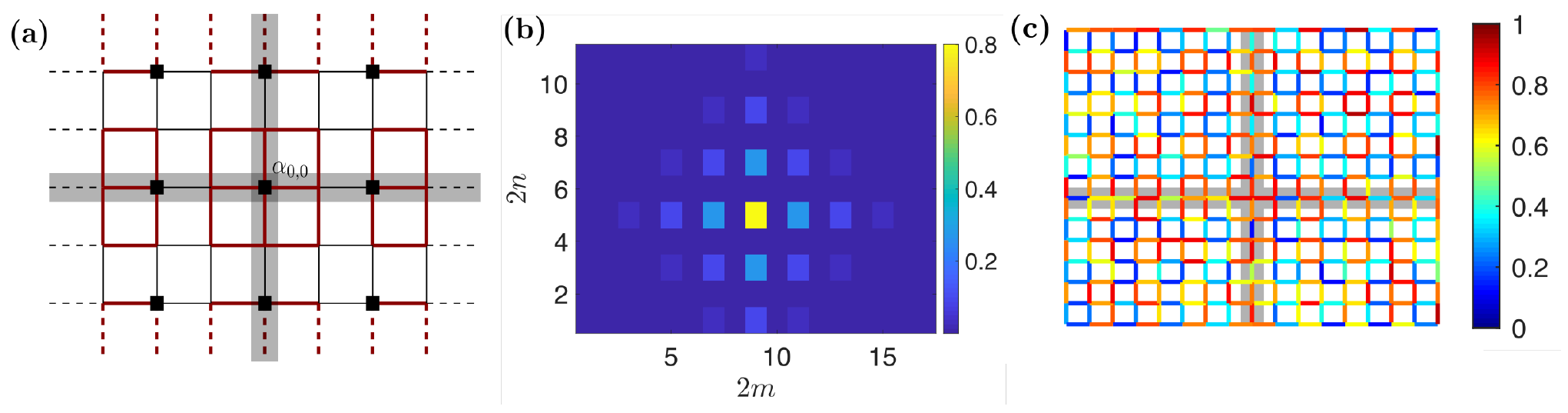} 
\caption{(a) Schematic construction of a defect-like mode in a network with zero fluxes (\eq{ZeroFlux_eq}). The hopping coefficients can be disordered, and the red and black indicate the one labeled by $s$ or $t$. Gray areas show the interfaces between the four quadrants. The defect mode is localized at the crossing (site $\alpha_{0,0}$) if $\langle \ln|s/t| \rangle < 0$. (b) Modulus of the components of a localized mode with $\varep=0$ in a network as (a) with zero flux disorder, $N_x=8$, $N_y=7$, $s_0 = 0.15$ and $t_0=0.85$. (c) Network displaying the defect mode of (b). Quadrants are delimited by the gray shading. 
}
\label{2D_DefectMode_Fig} 
\end{figure*}

Interestingly, all the results obtained above also apply to defect-like localized modes at the crossing between four networks with different topology. To see this, we can start again from the general zero energy solution of \eq{General_CornerMode} to build a defect-like localized solution in the middle of the network by choosing the appropriate average behavior of the hopping coefficients. To do so, we divide the network in four quadrants, and arrange $\langle \ln |s/t| \rangle$ for vertical and horizontal links to change sign in each quadrant such that the amplitude of \eq{General_CornerMode} decreases on average when moving away from the center, i.e. when $|m|$ and $|n|$ increase. The general construction is illustrated in Fig.~\ref{2D_DefectMode_Fig}(a), and an explicit example of such a network is shown in Fig.~\ref{2D_DefectMode_Fig}(c), which possesses the defect mode at $\varep=0$ shown in Fig.~\ref{2D_DefectMode_Fig}(b). 

%
%
\section{Acoustic realisation}
\label{Acoustic_Sec}
\begin{figure*}[t!]
\centering
\includegraphics[width=0.8\textwidth]{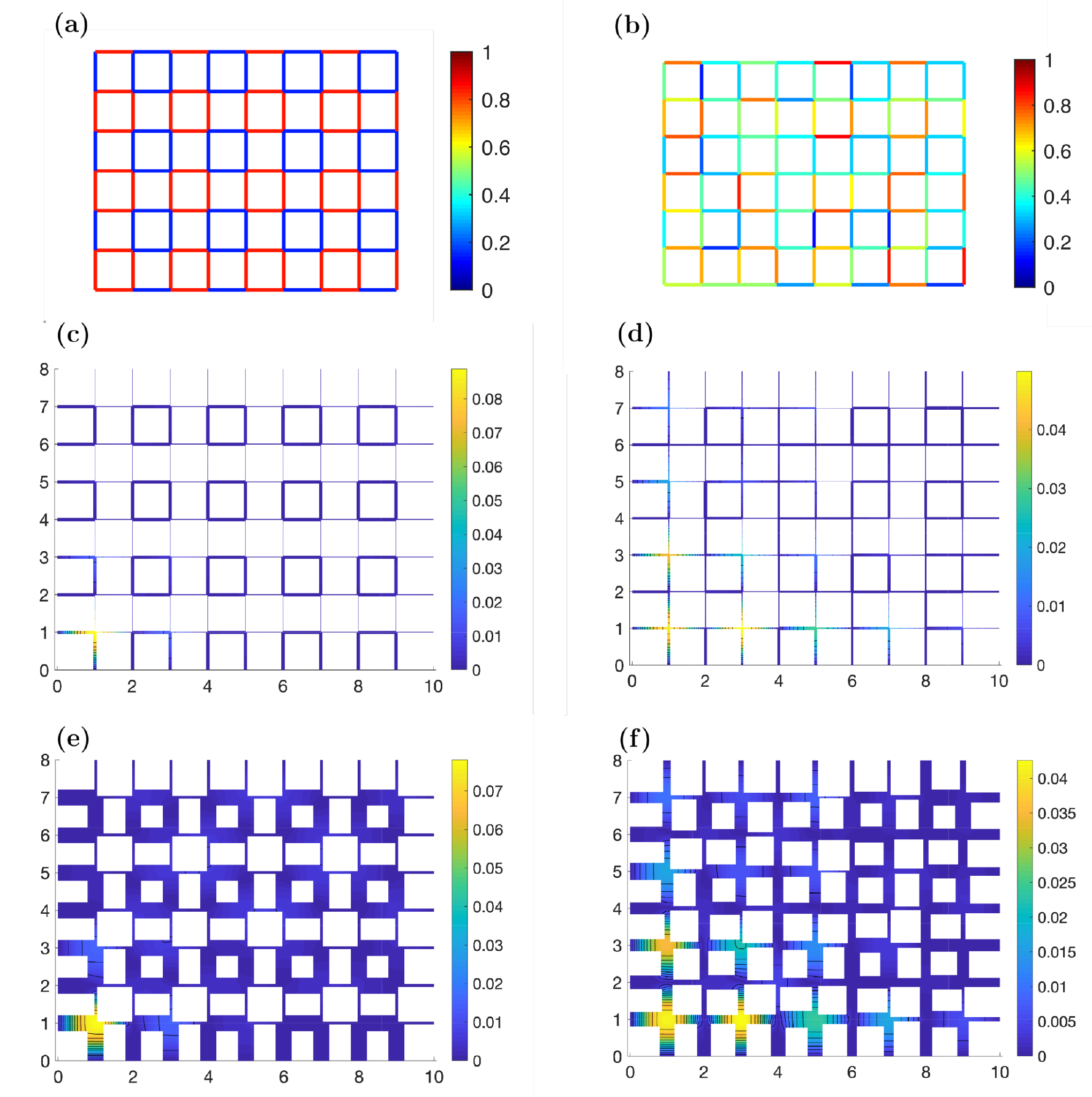} 
\caption{(a,b) Representation of 2D SSH asymmetried networks with color scale showing the hopping coefficients values. (a) Clean network. (b) Disordered network. (c,d,e,f) Corner mode in the two-dimensional acoustic realizations of the networks of (a,b). We used $N_x=4$, $N_y=3$, and cross-sections are such that $s_0=0.15$ and $t_0=0.85$. (c,d) $w/L = 0.015$ and $w'/L = 0.085$. (e,f) $w/L = 0.075$ and $w'/L = 0.425$. (c) The corner mode has $\varep_0 = -0.054$ and $\mathcal P_\alpha = 0.99$. (d) The corner mode has $\varep_0 = -0.068$ and $\mathcal P_\alpha = 0.99$. (e) The corner mode has $\varep_0 = -0.274$ and $\mathcal P_\alpha = 0.96$. (f) The corner mode has $\varep_0 = -0.337$ and $\mathcal P_\alpha = 0.99$. 
}
\label{2D_AcousticNet_Fig} 
\end{figure*}

In this section, we propose an acoustic realisation of the 2D SSH model with disorder. For this we consider a network of narrow air channels of equal length $L$ but varying cross-sections (see Fig~\ref{2D_AcousticNet_Fig}(c-f)). The typical transverse length $\ell_\perp$ of the channels is assumed much smaller that its length $L$ ($\ell_\perp \ll L$) so that inside each channel the propagation is monomodal~\cite{Depollier90,Wang17,Zheng20}. In~\cite{Zheng19} it was shown that using cross-sections alternating between two values $w$ and $w'$, the system is described by an effective Hamiltonian that coincides with the 2D SSH model with hopping coefficients 
\be \label{Clean_AcSSH}
s = \frac{w}{w+w'} \qquad {\rm and} \qquad t = \frac{w'}{w+w'}.
\ee
Here we show how this can be extended to a disordered 2D SSH model by using varying cross-sections for the channels. To see this, let us consider a node on the network, labeled by $a=(m,n,\mu)$ with $\mu \in \{\alpha, \beta, \gam, \delta \}$. The acoustic flux must be conserved at that node, implying 
\be \label{Cont_eq}
\sum_{\langle b,a \rangle} w_{ab} u_b = 0, 
\ee
where $u_b$ is the acoustic velocity arriving at the node $a$ from the channel connecting to a neighbouring node $b$ and $w_{ab}$ the section of that channel. Hence, the notation $\langle b,a \rangle$ means that the sum runs over all $b$ which are nearest neighbours to $a$. Moreover, pressure is continuous at the node $a$ and may be related to a neighbouring node $b$ by integrating the (1D) Helmholtz equation inside the channel. This leads to 
\be
\cos(kL) p_a + i \sin(kL) u_b = p_b,  
\ee
where $p_a$ and $p_b$ denote the acoustic pressure at nodes $a$ and $b$, and $k$ is the wavenumber. Summing over nearest neighbours $b$ and applying the debit continuity given by \eq{Cont_eq} then gives 
\be \label{Naive_Hamiltonian}
\varep p_a \sum_{\langle c,a \rangle} w_{ac} = 2\sum_{\langle b,a \rangle} w_{ab} p_b, 
\ee
where $\varep = 2\cos(k L)$ and the factor 2 is here to recover the same model as in preceding sections. In order to recast \eq{Naive_Hamiltonian} as a Hermitian eigenvalue problem for $\varep$, and apply the results of the preceding sections, we define the field $\Phi$ as rescaled pressure values: 
\bea
\phi_a &=& p_a \sqrt{\sum_{\langle c,a \rangle} w_{ac}} . 
\eea
Equation~\eqref{Naive_Hamiltonian} then rewrites as $\varep \Phi = H \cdot \Phi$, with the matrix elements of the (hermitian) Hamiltonian 
\be \label{Disorder_Ac_Coef}
H_{ab} = \frac{2w_{ab}}{\sqrt{\sum_{\langle c,a \rangle} w_{ac} \sum_{\langle c',b \rangle} w_{bc'}}}, 
\ee
for $a$ and $b$ nearest neighbours (and $0$ otherwise). For a finite network, the open boundary conditions of the 2D SSH model are obtain by adding extra channels with open ends. At the open ends, the acoustic pressure is at equilibrium with the exterior, and hence, must vanish~\footnote{We neglect radiative losses at the open ends, which vanish in the limit of small cross-sections.}. This means that the open ends act as ghost sites where the field amplitudes vanish, reproducing the open boundary conditions (see footnote~\ref{BC_ftn}). Hence, we have shown that the acoustic network in the limit $\ell_\perp \ll L$ is exactly described by the 2D SSH Hamiltonian of \eq{2D_SSH_Disorder}. This guarantees that acoustic networks in that limit possess the same properties such as edge waves, chiral symmetry, and corner modes and their robustness against disorder as previously studied. 

We computed the spectrum of the Helmholtz equation in a two-dimensional asymmetrized acoustic network using a finite element method (solving the 2D Helmholtz equation, $\Delta p +k^2 p=0$, with rigid wall Neumann boundary conditions). For commodity, we work in units where the length of each channel is unity ($L = 1$). In Fig.~\ref{2D_AcousticNet_Fig}(c-f) we show the obtained corner mode. We start with a clean network with thin channels with transverse lengths $w/L = 0.015$ and $w'/L = 0.085$ in Fig.~\ref{2D_AcousticNet_Fig}(c) and the corresponding disordered network with zero fluxes in Fig.~\ref{2D_AcousticNet_Fig}(d) (disorder construction is detailed in appendix~\ref{Disorder_App}). As predicted by the discrete model of the previous section, the corner mode is well localized and robust to this type of disorder. We notice that two-dimensional effects introduce a small breaking of the chiral symmetry, which manifests itself as a non-zero effective energy of the corner mode, with $\varep_0 = -0.054$ for the clean case (Fig.~\ref{2D_AcousticNet_Fig}(c)) and $\varep_0 = -0.068$ for the disordered case (Fig.~\ref{2D_AcousticNet_Fig}(d)). We also verify that the mode has support mostly on $\alpha$ sites. This is quantified using the sublattice polarization $\mathcal P_\alpha = \sum_{m,n} |p_\alpha^{m,n}|^2$ (with normalization $\sum_{\mu, m,n} |p_\mu^{m,n}|^2 = 1$), which is above $0.99$ for both clean and disordered networks. We also compute the corner mode for a network with transverse length five times larger. We show the results both in the clean case, Fig.~\ref{2D_AcousticNet_Fig}(e), and disordered case, Fig.~\ref{2D_AcousticNet_Fig}(f). Although two dimensional effects are more significant, with an effective energy $\varep_0 = -0.274$ for the clean case (Fig.~\ref{2D_AcousticNet_Fig}(e)) and $\varep_0 = -0.337$ for the disordered case (Fig.~\ref{2D_AcousticNet_Fig}(f)), the corner mode stays well localized on the $\alpha$-sites, with $\mathcal P_\alpha > 0.95$. Lastly, to make closer contact with the previous sections, we show the corresponding discrete networks in Fig.~\ref{2D_AcousticNet_Fig}(a,b).

%
%
\section{Conclusion}

In this work we study a two dimensional extension of the SSH model. This model has been shown to be a higher order topological insulator~\cite{Xie18,Ota19,Zhu20,Xu20}, hosting localized modes at corners in its topological phase. In this model, the corner modes have the peculiar property of coexisting with bulk waves as bound states embedded in the continuum (BIC). We study the robustness of these corner modes to the introduction of disorder that preserves the chiral symmetry of the model. We show that, while localization is rapidly lost in a general disorder, the corner modes are preserved up to high disorder strengths if the disorder satisfies the constraint of having zero fluxes through appropriately defined super plaquettes (see Fig.~\ref{2D_CornerPaths_Fig}). We also show that this condition can be seen at the level of the mode itself. We show that it is equivalent to the corner having support on a single site per unit cell, as in the clean case, while the chiral symmetry would only guarantee support of two sites per cell (see \eq{General_CornerMode}). 

This robustness goes against the intuition about BICs, where they are expected to lose their localization properties by hybridizing with bulk waves as soon as separability is broken. This was already noticed in periodic networks where separability was broken by extra couplings between next-to-nearest neighbour sites~\cite{Benalcazar20,Cerjan20}. In these works, it was however shown that robustness requires the presence of a crystalline symmetry ($C_{4v}$) on top of the chiral symmetry. Our results strengthen this conclusion and extends it to disordered networks, where the crystalline symmetry requirement is replaced by the condition of vanishing fluxes on super plaquettes. It should also be noticed that this condition of vanishing fluxes is rather mild, as can be seen for instance by counting the number of independent parameters compared to a general disorder. 

In the last section, we show how this model can be realized in an acoustic network made of air channels arranged in a square lattice. By varying the cross-sections of each channel, the disordered model is realized, with or without the zero-flux constraint. In the latter case, we confirm the presence of well localized corner modes by finite element simulations of the full network (see Fig.~\ref{2D_AcousticNet_Fig}). These results open the door to further refined manipulations of sound waves using higher order topological insulators by having localized modes even in the absence of a full bandgap.

\begin{acknowledgments}
This project has received funding from the European Union's Horizon 2020 research and innovation programme under the Marie Sklodowska-Curie grant agreement No 843152. 
\end{acknowledgments}

%
%
\appendix
\section{Few results for the 1D SSH chain}
\label{1D_App}
\begin{figure}[htp]
\centering
\includegraphics[width=0.9\columnwidth]{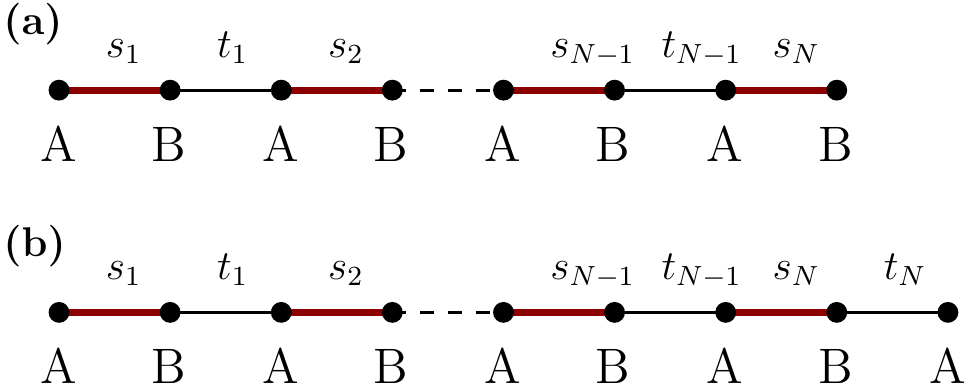}
\includegraphics[width=0.9\columnwidth]{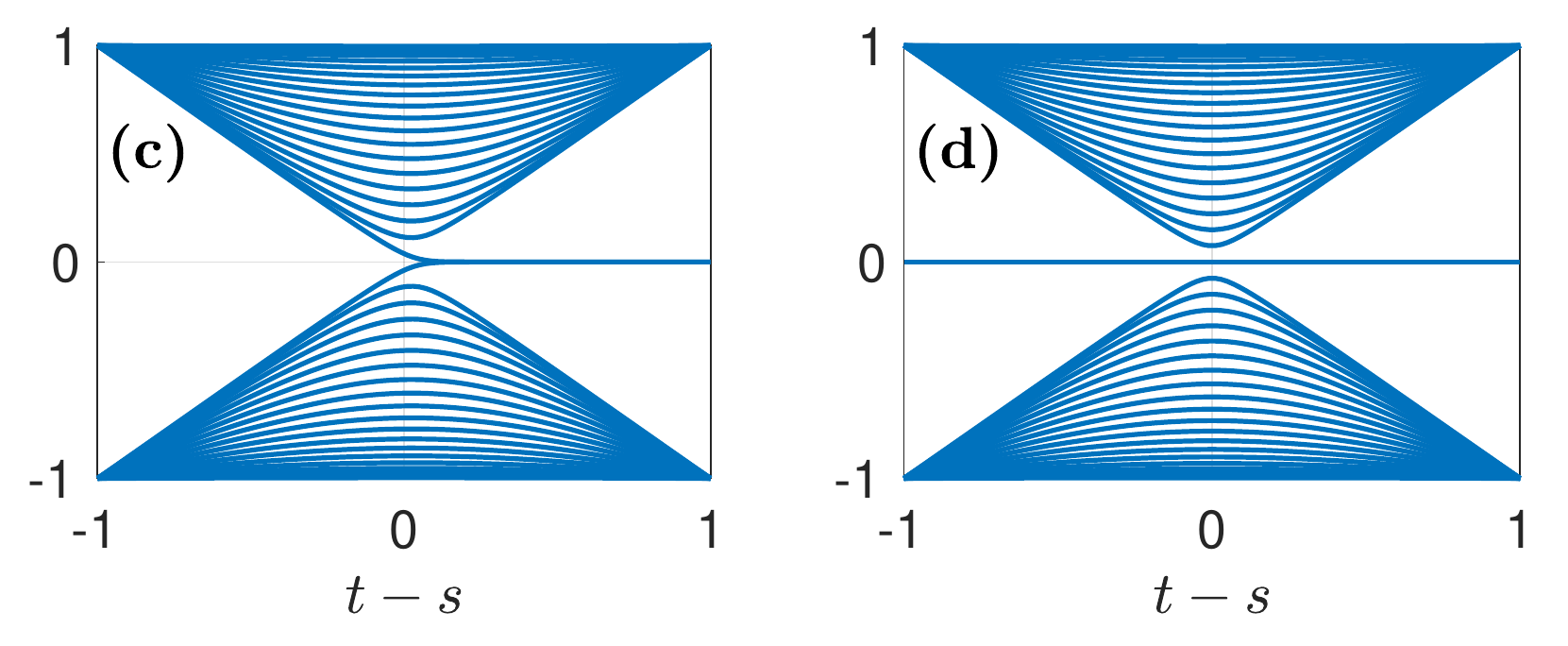}
\caption{(a-b) Representation of finite one-dimensional SSH chain. We also show the labels used for disordered chain, and used in \eq{SeparableDis_eq}. (c-d) Energy eigenvalues of a 1D chain ($N=20$) as a function of $t-s$. (a,c) Canonical chain, (b,d) Asymmetrized chain. 
}
\label{1DdisorderSSH_Schema_Fig} 
\end{figure}

In this appendix, we recall a few basic properties of the SSH model (see e.g.~\cite{Asboth16} for more details). The model is illustrated in Fig.~\ref{1DdisorderSSH_Schema_Fig}, and we assume the same constraint on the hopping coefficient as in the 2D case: $s>0$, $t>0$, and $s+t=1$. We first consider an infinite network, where the eigenmodes can be given in terms of Bloch waves $\Phi^m = e^{imq} \Phi$ (with $\Phi = (\phi_A, \phi_B)^T$) obeying $\varep \Phi = h(q) \cdot \Phi$ with the Bloch Hamiltonian 
\be
h(q) = \bmat 0 & s + t e^{-iq} \\ s + t e^{iq} & 0 \emat. 
\ee
The eigenvalues of $h(q)$ gives us the dispersion relation 
\be \label{1D_SSH_DispRel}
\varep^2 = |s + t e^{iq}|^2 = s^2 + t^2 + 2 st \cos(q), 
\ee
where we see the two bands of the model: $\varep \in [|t-s|, 1]$ and $\varep \in [-1, -|t-s|]$. In one dimension it is also rather easy to obtain the density of states as $\rho_{1D}(\varep) = 1/(2\pi \p_q \varep)$, hence  
\be \label{1D_DoS}
\rho_{1D}(\varep) = \frac{|\varep|}{\pi \sqrt{(1-\varep^2)(\varep^2 - \Delta^2)}}, 
\ee
for $(t-s)^2 < \varep^2 < 1$ and zero elsewhere. In appendix~\ref{2D_DoS_App} we show how to extend this result to the 2D case. 

Now, as discussed at the beginning of section~\ref{Finite_Net_Sec} in 2D, there are two ways to obtain a finite SSH chain: either taking an integer number $N$ of unit cells, as in Fig.~\ref{1DdisorderSSH_Schema_Fig}(a), or adding an extra site at the end, as in Fig.~\ref{1DdisorderSSH_Schema_Fig}(b). We first look for bulk wave solutions. They can be conveniently written as a superposition of left and right moving Bloch waves, that is 
\be
\Phi^m = \lam_1 e^{i m q} \bmat \varep \\ s+te^{iq} \emat + \lam_2 e^{-i m q} \bmat \varep \\ s+te^{-iq} \emat , 
\ee
where $\varep$ and $q$ are related by the dispersion relation \eqref{1D_SSH_DispRel}. Since changing $q$ into $-q$ leads to the same global solution, we can restrict ourselves to $q>0$. Using the fact that the chain is finite gives us two boundary conditions (which amount to adding an extra site on the left/right where amplitude is zero). The one on the left gives $\lam_1 + \lam_2=0$, and the one on the right gives the quantization conditions 
\be \label{Qcond_Canonical}
\sin \left(N q\right) + \frac{s}{t} \sin \left((N+1)q\right) = 0, 
\ee
for a canonical chain (Fig.~\ref{1DdisorderSSH_Schema_Fig}(a)), and 
\be \label{Qcond_Asym}
\sin \left((N+1)q \right) = 0 , 
\ee
for an asymmetrized chain (Fig.~\ref{1DdisorderSSH_Schema_Fig}(b)). Interestingly, for the latter the quantization condition, \eq{Qcond_Asym} has a simple set of solutions 
\be \label{Qcond_Bloch_Asym}
q_j = \frac{j\pi}{N+1}, 
\ee
with $j=1..N$, and the corresponding energy eigenvalues $\varep = \pm |s + t e^{iq_j}|$, while for a canonical chain, \eq{Qcond_Canonical} has no closed form solution~\footnote{We believe that this corrects a typographical error in~\cite{Obana19}, where the condition of \eq{Qcond_Bloch_Asym} was incorrectly used for a 2D SSH ribbon with the corresponding transverse chain having a canonical structure (as in Fig.~\ref{1DdisorderSSH_Schema_Fig}(a)). Separability together with equations~\eqref{Qcond_Canonical} and \eqref{Qcond_Asym} show that this is the case only if one add an extra chain on one side, similarly to Fig.~\ref{2D_SSH_Finite_Fig}(b).}. Similarly, edge states are easier to obtain for an asymmetrized chain (Fig.~\ref{1DdisorderSSH_Schema_Fig}(b)). Indeed, looking at a zero energy solution, we see that 
\be \label{1D_SimpleEdgeWave}
\Phi^m = \lam_0 \bmat 1 \\ 0 \emat (-s/t)^m, 
\ee
with $\lam_0$ a normalization constant, satisfies both boundary conditions: it vanishes on all $B$-sites, and hence on the ghost ones at both ends of the chain. If $s<t$ it is localized on the left edge, and if $s>t$ it is localized on the right edge. Moreover, when $s<t$, we see that this is also the solution of the semi-infinite chain obtained by sending the right boundary to infinity. When adding disorder (see Fig.~\ref{1DdisorderSSH_Schema_Fig}(b)), this expression becomes
\be \label{1D_DisorderEdgeWave}
\Phi^m = \lam_0 \bmat 1 \\ 0 \emat \prod_{j=1}^m (-s_j/t_j).  
\ee
For a clean canonical chain, \eq{1D_SimpleEdgeWave} only gives an approximate solution of an edge mode for $s<t$, with a second one on the other side and vanishing on $A$-sites. Eigenmodes are obtained as symmetric and anti-symmetric combination of the two edge modes, and have nonzero exponentially small energies $\pm \varep_0$ due to evanescent coupling~\cite{Asboth16}.  

\section{Few results for 2D SSH networks}
\subsection{Eigenstate density}
\label{2D_DoS_App}

\begin{figure}[htp]
\centering
\includegraphics[width=0.9\columnwidth]{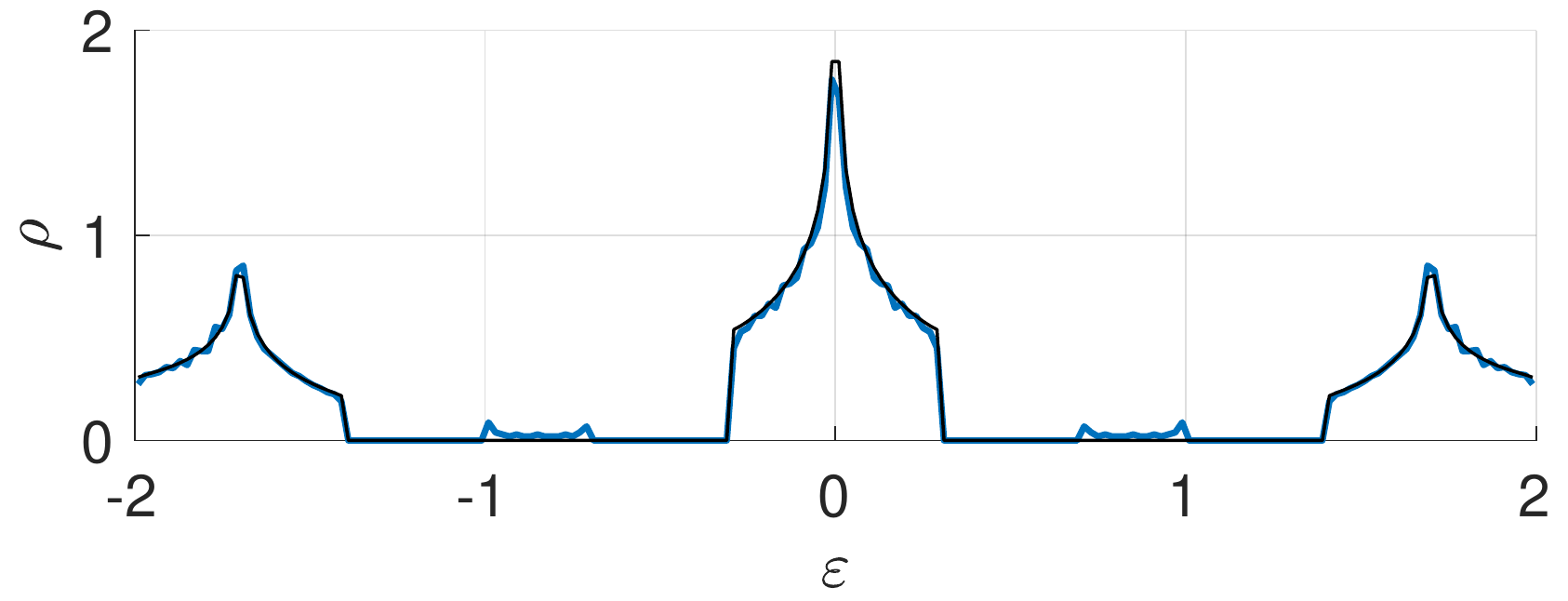}
\caption{Density of states $\rho$ computed numerically (blue) and analytically from \eq{2DSSH_DoS} (black) for $\Delta = 0.7$, $N_x = N_y = 50$. We notice the small non-zero density inside each gap: these are the edge waves, not included in the expression \eqref{2DSSH_DoS}. 
}
\label{2DSSH_Spectrum_Fig} 
\end{figure}

In this section we show that the separability of the 2D SSH model allows us to explicitly derive various useful quantities. For instance, in an infinite network, the density of states $\rho(\varep)$ can be obtained from its one-dimensional counterpart of \eq{1D_DoS}. For this, we write the density of states as $\rho(\varep) = \sum \int \delta(\varep - \varep(\mathbf q)) \d q_x \d q_y/(4\pi^2)$. Using the fact that $\varep(\mathbf q) = \varep_x(q_x) + \varep_y(q_y)$, and the change of variables from $q_{x/y}$ to $\varep_{x/y}$, which involves the 1D state density of \eq{1D_DoS}, we obtain 
\be \label{2DSSH_DoS}
\rho(\varep) = \int_{-1}^1 \rho_{1D} \left(\frac{\varep}{2} + \xi \right) \rho_{1D} \left(\frac{\varep}{2} - \xi \right) \d \xi. 
\ee
In Fig.~\ref{2DSSH_Spectrum_Fig} we compare this expression with a direct numerical computation of the density of states. We see that the density of states becomes very large near the middle of each band, which correspond to logarithmic singularities. To understand this, the first step is to notice that $\rho_{1D}$ has a square root singularity near each band edge, which is integrable. Hence, the integrant in \eq{2DSSH_DoS} is singular if both $\varep/2 + \xi$ and $\varep/2 - \xi$ are at a singular point of $\rho_{1D}$. Moreover, because the singularities of $\rho_{1D}$ are one-sided, these two points must correspond to different sides. This happens for $(\varep/2 + \xi,\varep/2 - \xi)$ equal to $(-1,-\Delta)$, $(-\Delta,\Delta)$, $(\Delta,1)$, or $(-1,1)$, which leads to $(\varep, |\xi|)$ being $(-(1+\Delta),(1-\Delta)/2)$, $(0,\Delta)$, $(1+\Delta , (1-\Delta)/2)$, or $(0, 1)$, so indeed when $\varep$ is at the center of a band. Because the singularity of the integrant is $O(1/\xi)$, we can anticipate a logarithmic divergence of the density of states. Let us show this for instance near $\varep = 0$. Using the parity of $\rho$, and denoting the vicinity of a point by $\mathcal V$, we write the integral representation \eqref{2DSSH_DoS} as 
\bea 
\rho(\varep) &\sim& 2\int_{\mathcal V(\Delta) \cup \mathcal V(1)} \rho_{1D} \left(\frac{\varep}{2} + \xi \right) \rho_{1D} \left(\frac{\varep}{2} - \xi \right) \d \xi , \nonumber \\
&\sim& \frac{|\Delta|}{\pi^2 (1-\Delta^2)} \int_{\mathcal V(0)} \frac{\d \zeta}{\sqrt{\zeta^2 - \varep^2/4}} \nonumber \\
&& + \frac{1}{\pi^2 (1-\Delta^2)} \int_{\mathcal V(0)} \frac{\d \zeta}{\sqrt{\zeta^2 - \varep^2/4}} , 
\eea
hence, 
\bea
\rho(\varep) &\sim& -\frac{1+|\Delta|}{\pi^2 (1-\Delta^2)} \ln\left|\frac{\varep}{2} \right|. 
\eea 

\subsection{Corner mode equation}
\label{CornerMode_App}

In the core of this work, we derived an explicit corner mode solution by assuming that the corner mode has support only on $\alpha$-sites. This leads to \eq{2D_DisorderSSH_WithCmode}. Here, we give extra details on how this equation is obtained. We start by writing the eigen-value equation $\varep \Phi = H \cdot \Phi$ with the disordered Hamiltonian of \eq{2D_SSH_Disorder} in components: 
\bsub \bea
\varep \phi_\alpha^{m,n} &=& s_{m,n}^{(1)} \phi_\beta^{m,n} + s_{m,n}^{(2)} \phi_\gamma^{m,n} \nonumber \\
&& + t_{m,n-1}^{(1)} \phi_\beta^{m,n-1} + t_{m-1,n}^{(2)} \phi_\gamma^{m-1,n} , \\
\varep \phi_\beta^{m,n} &=& s_{m,n}^{(1)} \phi_\alpha^{m,n} + s_{m,n}^{(4)} \phi_\delta^{m,n} \nonumber \\
&&+ t_{m,n}^{(1)} \phi_\alpha^{m,n+1} + t_{m-1,n}^{(4)} \phi_\delta^{m-1,n} , \\
\varep \phi_\gamma^{m,n} &=& s_{m,n}^{(2)} \phi_\alpha^{m,n} + s_{m,n}^{(3)} \phi_\delta^{m,n} \nonumber \\
&&+ t_{m,n}^{(2)} \phi_\alpha^{m+1,n} + t_{m,n-1}^{(3)} \phi_\delta^{m,n-1} , \\
\varep \phi_\delta^{m,n} &=& s_{m,n}^{(3)} \phi_\gamma^{m,n} + s_{m,n}^{(4)} \phi_\beta^{m,n} \nonumber \\
&&+ t_{m,n}^{(3)} \phi_\gamma^{m,n+1} + t_{m,n}^{(4)} \phi_\beta^{m+1,n} . 
\eea \esub
We now assume that the corner mode has support only on $\alpha$-sites, i.e. $\phi_\beta^{m,n} = \phi_\gamma^{m,n} = \phi_\delta^{m,n} = 0$. The previous set of equation becomes 
\bsub \bea
\varep \phi_\alpha^{m,n} &=& 0 , \\
0 &=& s_{m,n}^{(1)} \phi_\alpha^{m,n} + t_{m,n}^{(1)} \phi_\alpha^{m,n+1} , \\
0&=& s_{m,n}^{(2)} \phi_\alpha^{m,n} + t_{m,n}^{(2)} \phi_\alpha^{m+1,n} , \\
0 &=& 0. 
\eea \esub
This implies that the corner mode must have zero energy, and satisfy \eq{2D_DisorderSSH_WithCmode}. 

\subsection{Connection between the corner mode and biased random walks}
\label{2DRandomWalk_App}
In the one-dimensional SSH model with disordered hopping coefficients, it is known that the chiral symmetric point ($\varep = 0$) is governed by a random walk dynamics. This leads to the identification of several exotic properties, such as anomalous localization~\cite{Inui94} or density-of-state singularities~\cite{Theodorou76,Eggarter78}. In disordered 2D SSH networks with zero flux (\eq{ZeroFlux_eq}), the product state structure of the zero mode found in \eq{General_CornerMode} suggests that its properties can also be derived from random walk dynamics. More precisely, the governing equation of the zero mode in disorders with zero flux, namely \eq{2D_DisorderSSH_WithCmode}, can be recast as two independent random walks. To see this, let us solve \eq{2D_DisorderSSH_WithCmode} with separation of variable, writing $\phi_\alpha^{m,n} = \phi_{\rm x}^m \times \phi_{\rm y}^n$. We then consider the logarithm of the amplitude to obtain the two equations  
\bsub \bea
\ln\left| \phi_{\rm y}^{n+1} \right| &=&\ln\left| \phi_{\rm y}^{n} \right| + \ln\left| \frac{s_{m,n}^{(2)}}{t_{m,n}^{(2)}} \right| , \label{2D_RandWalk_y} \\
\ln\left| \phi_{\rm x}^{m+1} \right| &=&\ln\left| \phi_{\rm x}^{m} \right| + \ln\left| \frac{s_{m,n}^{(4)}}{t_{m,n}^{(4)}} \right| \label{2D_RandWalk_x} . 
\eea \esub
Technically, \eq{2D_RandWalk_y} (resp. \eq{2D_RandWalk_x}) depends on the value of $m$ (resp. $n$) along which one integrates. However, the condition of zero flux of \eq{ZeroFlux_eq} ensures that the result is consistent since the sum $\ln\left| \phi_{\rm x}^{m} \right|+\ln\left| \phi_{\rm y}^{n} \right|$ is independent of these choices. 

This connection to random walks allows us to easily obtain the zero-mode behavior, as discussed in section~\ref{AnalyticCorner_Sec}, after \eq{General_CornerMode}: if $\ln|s/t|$ has a non-zero value, the random walks drift in a preferred direction (balistic regime), meaning that the field amplitude grows or decays exponentially; if $\langle \ln|s/t| \rangle = 0$, then there is no preferred direction but the random walks diverge as $\propto \sqrt n$ (diffusive regime), leading to an anomalous localization of the field $\propto e^{-\lam \sqrt m - \lam' \sqrt n}$. As a last remark, we point out that these properties inherited from random walks are valid under the constraint of zero fluxes of \eq{ZeroFlux_eq}. Unconstrained but chiral disorders have dramatically different properties, see e.g.~\cite{Mudry03,Evangelou03} for the studies in the regime $\langle \ln|s/t| \rangle = 0$.

\subsection{Corner modes in disordered kagome lattices}
\label{Kagome_App}

\begin{figure}[htp]
\centering
\includegraphics[width=0.9\columnwidth]{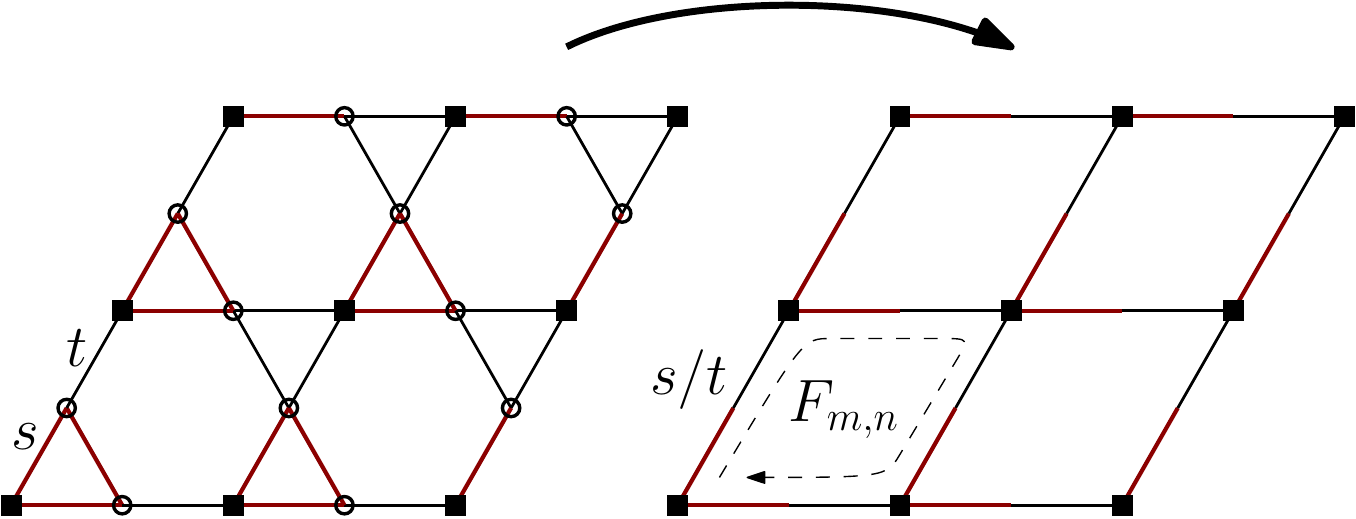}
\caption{Super plaquettes and corresponding flux in a kagome lattice, to compare with Fig.~\ref{2D_CornerPaths_Fig}(a). On the left, we marked with open circles the sites of the lattice where the corner mode (located on the lower left side) has zero amplitude. We also point out that the hopping coefficients for each link are a priori different (although we emphasized the $s$ and $t$ types with red and black). 
}
\label{Kagome_Fig} 
\end{figure}

Here we explain how the general construction of the corner mode \eqref{General_CornerMode} with the robutness condition, i.e. vanishing fluxes as in \eqref{ZeroFlux_eq}, can be applied to other situations. We already mentioned the relation to corner modes in Lieb lattices~\cite{Poli17}. Corner modes have also drawn significant interest in kagome lattices~\cite{Ni19,Xue19}. Although kagome lattices are not chiral, corner modes have a nontrivial sublattice structure, with two sites with vanishing amplitudes out of the three per unit cell. By removing the links that are irrelevant to the corner mode, because they connect sites with zero amplitude, we see that the same structure as in the 2D SSH model appears. Hence the same construction of the corner mode, and its robustness condition (zero fluxes \eqref{ZeroFlux_eq}). This is illustrated in Fig.~\ref{Kagome_Fig}, to compare with Fig.~\ref{2D_CornerPaths_Fig}(a).

We also notice that a similar general construction was obtained in periodic networks (where the fluxes of \eq{ZeroFlux_eq} are always trivial) in~\cite{Kunst19}.

\subsection{Constructing different types of disorder}
\label{Disorder_App}
In this appendix, we explain how to obtain the different types of disorder, by modifying \eq{2D_Disorder_H}. 
\bigskip

\noindent {\bf Separable disorders:} 
To obtain a separable disorder, we pick random hopping coefficients between 0 and 1 for two 1D chains, and obtain a disordered matrix $W_{\rm sep}$ using \eq{SeparableDis_eq}. We then build the Hamiltonian $H(\Delta)$ by interpolation between the clean network and that disorder, as in \eq{2D_Disorder_H}. 
\bigskip

\noindent {\bf Zero flux disorders:} 
To obtain the disorder with zero-fluxes, we start from a general disorder (\eq{2D_Disorder_H}) and compute the flux $F_{m,n}$: if it is larger than unity, we rescale $s_{m,n+1}^{(2)}$, if it is smaller than unity, we rescale $s_{m+1,n}^{(1)}$. Doing so, the corresponding hopping coefficients are lowered, and hence stay between 0 and 1. We also rescale $s$ rather than $t$ because lowering the former changes the statistical spread less~\footnote{Notice also that these rescalings are consistent by scanning the super plaquettes in ascending order so that preceding fluxes are unaffected when changing a given hopping coefficient.}. This procedure leads to a continuous family of Hamiltonians $H(\Delta)$ with a comparable disorder strength as the others for a given value of $\Delta$. 
\bigskip

\noindent {\bf Acoustic disorders with zero fluxes:} 
In the acoustic realization of section~\ref{Acoustic_Sec}, it is trickier to infer the disorder structure from that of the cross-section values. For instance, having the cross-sections depending only on one coordinate (horizontal or vertical) is not enough to obtain the separable disorder described in \eq{SeparableDis_eq}, because of the denominator of \eqref{Disorder_Ac_Coef} involving neighbouring cross-sections. We can however obtain disorders with zero fluxes, as in \eqref{ZeroFlux_eq}. To do so we start by randomly taking the cross-section values around a staggered mean and of spread $\Delta$, as in \eq{2D_Disorder_H}. We then compute the fluxes $F_{m,n}$, and rescale the appropriate cross-section values to obtain a vanishing flux. By doing so one super plaquette after another however, we also affect previously trivialised fluxes (again due to the denominator of \eqref{Disorder_Ac_Coef}). Hence we proceed iteratively: scanning through each super plaquette to trivialise the corresponding flux, and then restarting the procedure until all fluxes are zero. In practice after a few tries (about 10) the procedure stops and all fluxes are zero to numerical precision.

\subsection{Chiral and partial chiral symmetries}
\label{Chiral_App}
Just like its one-dimensional counterpart, the 2D SSH model is chiral symmetric. This means that there is a unitary operator $\Gamma$ that acts inside each cell (i.e. commutes with translations) such that $\Gam^2 = 1$ and 
\be \label{Chiral_eq}
\Gam \cdot H \cdot \Gam = - H.  
\ee 
In the 2D SSH model, we see that this is satisfied by defining  
\be
\Gam \cdot \bmat \phi_\alpha^{m,n} \\ \phi_\beta^{m,n} \\ \phi_\gamma^{m,n} \\ \phi_\delta^{m,n} \emat = \bmat -\phi_\alpha^{m,n} \\ \phi_\beta^{m,n} \\ \phi_\gamma^{m,n} \\ -\phi_\delta^{m,n} \emat . 
\ee
Hence, $\Gam$ leaves the $\beta$ and $\gamma$ sites (first sublattice) invariant and flips the sign on the $\alpha$ and $\delta$ sites (second sublattice). As a consequence, the spectrum is symmetric about 0: eigenvectors come in pairs $(\Phi, \Gam \cdot \Phi)$ associated with eigenvalues $(\varep, - \varep)$. Moreover, because it has a vanishing energy, a corner mode $\Phi_0$ is chiral invariant, i.e. $\Gam \cdot \Phi_0 = \pm \Phi_0$. In other words, $\Phi_0$ is guaranteed to vanish on one of the two sublattices. 

In the absence of disorder, the separability of the Hamiltonian into a product of 1D SSH chains leads to additional hidden chiral properties. Indeed, when writing the Hamiltonian as $H_0 = H_{0x} \otimes I_{2N_y+1} + I_{2N_x+1} \otimes H_{0y}$, each component $H_{0x}$ and $H_{0y}$ possesses its own chiral symmetry. Therefore, we define what we call ``partial chiral operators'' as the chiral operators associated with the corresponding 1D horizontal and vertical chains: 
\be 
\Gam_x \cdot \bmat \phi_\alpha^{m,n} \\ \phi_\beta^{m,n} \\ \phi_\gamma^{m,n} \\ \phi_\delta^{m,n} \emat = \bmat \phi_\alpha^{m,n} \\ -\phi_\beta^{m,n} \\ \phi_\gamma^{m,n} \\ -\phi_\delta^{m,n} \emat , 
\ee
and
\be 
\Gam_y \cdot \bmat \phi_\alpha^{m,n} \\ \phi_\beta^{m,n} \\ \phi_\gamma^{m,n} \\ \phi_\delta^{m,n} \emat = \bmat -\phi_\alpha^{m,n} \\ -\phi_\beta^{m,n} \\ \phi_\gamma^{m,n} \\ \phi_\delta^{m,n} \emat .
\ee
We see that $\Gam = \Gam_x \cdot \Gam_y$, which is why they are referred to as partial. $H_0$ is invariant under neither of these partial chiral operators. However, $H_{0x}$ (resp. $H_{0y}$) is chiral under $\Gam_x$ (resp. $\Gam_y$), while it commutes with $\Gam_y$ (resp. $\Gam_x$). As a consequence, each bulk eigen-vector of $H$, written as a product $\Phi = \psi \otimes \varphi$ is associated with three other bulk eigen-vectors $\Gam_x \cdot \psi \otimes \varphi$, $\psi \otimes \Gam_y \cdot \varphi$, and $\Gam \cdot \Phi = \Gam_x \cdot \psi \otimes \Gam_y \cdot \varphi$~\footnote{Notice that by construction $\Gam_x$ and $\Gam_y$ act separately on a tensor product: $\Gam_x \cdot (\psi \otimes \varphi) = (\Gam_x \cdot \psi) \otimes \varphi$ and $\Gam_y \cdot (\psi \otimes \varphi) = \psi \otimes (\Gam_y \cdot \varphi)$.}. Similarly, edge waves are invariant under one of the partial chiral operators and paired with another edge wave using the other partial chiral operator. 

Lastly, the corner mode is invariant under both partial chiral operators, i.e. $\Gam_x \cdot \Phi_0 = \pm \Phi_0$ and $\Gam_y \cdot \Phi_0 = \pm \Phi_0$. This last point is crucial to the present discussion, as it implies that the corner mode vanishes on three sites per cell, while chiral symmetry itself only guarantees it to vanish on two sites per cell. As we saw in section~\ref{AnalyticCorner_Sec}, this peculiar sublattice structure is at the origin of the robustness against disorder.

\bibliographystyle{utphys} 
\bibliography{Bibli}

\end{document}